\documentclass{jpp}
\usepackage{graphicx}

\usepackage[utf8]{inputenc}
\usepackage[T1]{fontenc}
\usepackage{amsmath}
\usepackage{xcolor}%
\usepackage{appendix}

\newcommand{\beq}{\begin{equation}}
\newcommand{\eeq}{\end{equation}}

\newcommand{\vR}{\boldsymbol{R}}
\newcommand{\vE}{\boldsymbol{E}}
\newcommand{\vB}{\boldsymbol{B}}
\newcommand{\vx}{\boldsymbol{x}}
\newcommand{\vy}{\boldsymbol{y}}
\newcommand{\vz}{\boldsymbol{z}}

\newcommand{\hz}{\hat{\boldsymbol{z}}}
\newcommand{\hG}{\hat{\Gamma}_0}
\newcommand{\vth}{v_{\rm th}}
\newcommand{\vparin}{v_{\parallel 0}}
\newcommand{\vperpin}{v_{\perp 0}}
\newcommand{\vph}{v_{\mathrm{ph}}}
\newcommand{\vA}{v_{\mathrm{A}}}
\newcommand{\rhoth}{\rho_{\rm th}}

\newcommand{\vref}{v_{\rm ref}}
\newcommand{\dd}{\mathrm{d}}

\newcommand{\kgk}[1]{{#1}} 

\newcommand{\am}[1]{{#1}}
\newcommand{\markup}[1]{{\bf{#1}}}

\shorttitle{Magnetic moment breaking by coherent fluctuations}
\shortauthor{A. Mallet}

\title{Perpendicular ion heating in turbulence and reconnection: magnetic moment breaking by coherent fluctuations}

\author{Alfred Mallet\aff{1}
  \corresp{\email{alfred.mallet@berkeley.edu}}, 
  Kristopher G. Klein\aff{2}, 
  Benjamin D.G. Chandran\aff{3}, Tamar Ervin\aff{1, 4}, Trevor A. Bowen\aff{1}}

\affiliation{\aff{1}Space Sciences Laboratory, University of California, Berkeley CA 94720, USA
\aff{2}Lunar and Planetary Laboratory, University of Arizona, Tucson, AZ 85719, USA
\aff{3}Space Science Center and Department of Physics, University of New Hampshire, Durham, NH 03824, USA
\aff{4}Department of Physics, University of California, Berkeley CA 94720, USA
}
\begin{document}

\maketitle
\begin{abstract}
We study the interaction of an ion with a fluctuation in the electromagnetic fields that is localized in both space and time. 
We study the scale-dependence of the interaction in both space and time, deriving a generic form for the ion's energy change, which involves an exponential cutoff based on the characteristic timescale of the electromagnetic fluctuation. 
This leads to diffusion in energy in both $v_\perp$ and $v_\parallel$. 
We show how to apply our results to general plasma physics phenomena, and specifically to Alfv\'enic turbulence and to reconnection. 
Our theory can be viewed as a unification of previous models of stochastic ion heating, cyclotron heating, and reconnection heating in a single theoretical framework.
\end{abstract}

\section{Introduction}
In both the solar corona and solar wind, observations show that proton heating is typically much greater than the electron heating, with minor ions heated even more strongly, and moreover that ion heating is mainly perpendicular to the magnetic field \citep{kohl1998,antonucci2000,marsch1982,marsch2004,hellinger2006,Kasper:2017,bowen2020}. 
\kgk{Characterizing} ion heating is therefore essential for the thermodynamics of \am{this system} \citep{parker1965}. 
\am{More generally, correctly parametrizing the ratio of \kgk{ion-to-electron} heating in plasma turbulence is of great interest for the interpretation of many remote astrophysical observations \citep{chael2018}.}

What is the source of free energy for \am{the observed} heating? 
One successful model is heating from the Alfv\'enic plasma turbulence ubiquitous in the solar wind \citep{belcher1971,chen2016,chen2020} and corona \citep{depontieu2007}: the fluctuation amplitudes are consistent with the observed plasma heating \citep{chandran2009,cranmer2009}, suggesting that the solar wind is accelerated and locally heated by the dissipation of \kgk{these turbulent fluctuations}. 
It is worth noting that, because the turbulence has only a very small compressive component \citep{klein2012}, the ion heating within the gyrokinetic approximation would be much small to explain the observations \citep{schektome2009,schekochihin2019,kawazura2020}.

Several theoretical models have been proposed to explain ion heating in turbulence. 
First, cyclotron resonant heating \citep{hollweg2002,chandran2010,isenberg2011,isenberg2019,bowen2024} occurs when, in the frame moving with an ion's parallel velocity $v_\parallel$, a wave's frequency $\omega - k_\parallel v_\parallel$ matches the gyrofrequency $\Omega_i$: hence "resonant", $\omega - k_\parallel v_\parallel - n \Omega_i =0$. 
This is often discussed in the framework of quasilinear theory \citep{kennel1966,stix1992}, where the resulting diffusion of energy in phase space is derived assuming \am{a spectrum of} infinite plane waves and considering only the resonant response. 

Another important model, closer in approach to that of the present paper, is stochastic heating \citep{mcchesney1987,chandran2010}, in which ions random-walk in energy due to uncorrelated kicks from ion-scale fluctuations. 
In the \citet{chandran2010} model, this results in a heating rate $Q_\perp\sim \delta u_\rho^3/\rhoth \exp(-c_2 \vth/\delta u_\rho)$, where $\delta u_\rho$ is the amplitude of $E\times B$ velocity fluctuations at the gyroscale $\rhoth$, and the exponential suppression factor was added empirically to account for the near-conservation of the magnetic moment at low frequencies \am{and small amplitudes.} 
One advantage of stochastic heating as opposed to cyclotron resonant heating is that it does not require an exact resonance, nor does it assume that the fluctuations resemble infinite plane waves. 
This allows one to easily incorporate the observed intermittent probability distribution of fluctuation amplitudes \citep{Chandran14,ms16} at the gyroscale, which can dramatically increase the predicted heating rate \citep{mallet2019,cerri2021}.

Observations \citep{chenmallet} show that the solar wind turbulence is highly anisotropic: fluctuations have very different characteristic lengthscales parallel ($l_\parallel$) and perpendicular ($\lambda$) to the background magnetic field, $l_\parallel \gg l_\perp$. 
Modern turbulence theories \citep{gs95,boldyrev} explain this in terms of a critical balance between characteristic timescales associated with linear propagation ($\tau_{lin}\sim l_\parallel / \vA$, with $\vA=B_0/\sqrt{4\pi n_p m_p}$ the Alfv\'en velocity based on the mean magnetic field $B_0$) and nonlinear interactions ($\tau_{nl}\sim \lambda / \delta u_\lambda$): the cascade time $\tau\sim\tau_{lin}\sim \tau_{nl}$, whence $l_\parallel / \lambda \sim \vA/\delta u_\lambda\gg1$. 
Since the fluctuating field amplitude $\delta u_\lambda$ at scale $\lambda$ is an increasing function of $\lambda$, at progressively smaller scales, the anisotropy $l_\parallel/\lambda$ increases. 
This means that the fluctuations remain relatively low-frequency, with $\omega \sim 1/\tau\ll \Omega_i$, where $\Omega_i=Z_i e B/m_i c$ is the ion gyrofrequency. 
\am{This poses a challenge: the magnetic moment $\mu = m_i v_\perp^2/B$ is conserved to all orders in $\eta \sim\omega/\Omega_i\ll1$ \citep{kruskal1962}, and so the usual perturbation theory would suggest that perpendicular ion heating should be irrelevant for such anisotropic, small-amplitude turbulence, in contrast to the observations.} 
It is worth noting that "to all orders" is not the same as "exactly": as an example (that will be important in this paper), $\exp(-1/\eta)\neq0$, but is "zero to all orders" if $\eta\ll1$, since all derivatives vanish as $\eta\to0$. 

Besides turbulence, magnetic reconnection has been proposed as a mechanism for coronal heating \citep{klimchuk2015} and also as a heating mechanism within the turbulence itself \citep{shay2018}. 
In fact, turbulent heating and reconnection heating may not be as distinct as traditionally thought. 
The turbulent cascade naturally leads to the formation of extended current sheets \citep{boldyrev,Chandran14,ms16}, which reconnect once their width becomes sufficiently small \citep{msc_disruption,msc_cless,bl2017,loureiroboldyrev,loureiroboldyrev_cless,vech2018,comisso2017,cerri2017,franci2017,dong2022}. 
Similarly, approaching the problem from the "reconnection end", extended reconnecting current sheets are often violently unstable, leading naturally to strong turbulence \citep{loureiro2007,bhattacharjee2009,huang2016}. 
Seeking to explain preferential heavy ion heating in the corona and solar flares, \citet{drake2009b} developed a theory of perpendicular ion heating in reconnection exhausts, supported by numerical simulations. 
In \citet{drake2009a}, they showed that in guide-field reconnection, strong ion heating only occurs if the characteristic timescale to transit the exhaust is shorter than the ion's gyroperiod. 
This behaviour has similarities to the stochastic heating in turbulence.

All three of cyclotron-resonant, stochastic, and reconnection perpendicular ion heating share a common feature: they require the conservation of the magnetic moment to be broken. 
In the case of stochastic and reconnection heating, this leads to a "threshold" which must be satisfied for strong ion heating to be possible. 
Likewise, in cyclotron resonant heating, the resonance condition must be satisfied (we will argue that this "sharper" behaviour is a consequence of the plane-wave assumption). 
\citet{johnston2025} noticed the similarities between cyclotron-resonant and stochastic heating, and found that the heating in their test-particle simulations was well described by a single exponential suppression factor modelling both cyclotron-resonant heating in imbalanced turbulence and stochastic heating in balanced turbulence. In related work in a different physical setting, namely electron scattering in the radiation belts, a similarly close relationship between electron scattering by strongly nonlinear coherent structures (electron holes) and quasilinear theory was also previously discussed by \citet{vasko2017} and \citet{vasko2018}.

In this paper, we develop a new framework that describes perpendicular ion heating. 
We analytically study the response of an ion to a localized, coherent fluctuation in the electromagnetic fields, with the fluctuating electric $\delta\vE$ and magnetic $\delta\vB$ fields tending to zero at $t=\pm\infty$, an approach to our knowledge first taken in \citet{krall1964} for electric field fluctuations and for general adiabatic invariants in \citet{landau1976}. 
Quite generically, we find that the perpendicular ion kinetic energy $m_i v_\perp^2/2$ changes by an amount of order
\begin{equation}
    m_i \Delta\sim m_i\epsilon \exp(-1/\eta),\label{eq:echange}
\end{equation}
where $\Delta$ is the change in $v_\perp^2$, $\epsilon\sim \delta B/B_0 \sim c\delta E/B_0v_\perp$ is the normalized amplitude of the fluctuations and $\eta\sim 1/\tau\Omega_i$, where $\tau$ is a characteristic timescale over which $\delta \vB$ and $\delta \vE$ vary. 
The threshold for strong ion heating to occur is encoded in the exponential factor: $\mu$-conservation is lost when $\eta\sim1$ and the fluctuations vary significantly over one gyroperiod. 
\am{For $\eta\ll1$, the magnetic moment is conserved to all orders, but not exactly: for many systems, this is enough to provide significant heating over long timescales.}
After setting up the system of equations (Sec.~\ref{sec:norm}), \am{ in Sec.~\ref{sec:main} we proceed to expand in the amplitude of the fluctuating fields, deriving analytic expressions for the change in perpendicular and parallel energy as well as how this depends on the lengthscale of the fluctuations. 
We also derive general formulae for the diffusion coefficient and heating rate, and outline how our theory should be applied to different physical systems. }
We then explicitly show how our results apply to both Alfv\'enic turbulence (Sec.~\ref{sec:turb}) and to reconnection (Sec.~\ref{sec:rx}). 
Finally, we discuss the relationship of our model to earlier theories, and what the implications of our results are for astrophysical and space plasma turbulence and reconnection heating.

\section{Normalized equations}\label{sec:norm}
The equations of motion for an ion of charge $Z_i e$ and mass $m_i$ in a general electromagnetic field are
\am{\begin{equation}
    \frac{\mathrm{d}^2 \vR}{\mathrm{d}t^2} = \frac{Z_ie}{m_i}\left[\vE + \frac{\mathrm{d}{\vR}/\mathrm{d} t\times \vB}{c}\right].
\end{equation}
For the magnetic and electric field, we take }
\begin{equation}
    \vB = B_0\left(\hat{\vz} + \varepsilon b(y,z,\Omega_i t)\hat{\vx}\right) , \quad \vE = \frac{\varepsilon \vperpin B_0}{c}g(y,z,\Omega_it)\hat{\vy}
\end{equation}
where we assume $\varepsilon\ll1$, and $\vperpin$ is the perpendicular ion velocity at $t=-\infty$. 
Our neglect of $B_y$ and $E_x$ will not change the physical conclusions of our calculation (while making it somewhat less cumbersome), but ignoring $E_z$ and fluctuations in $B_z$ removes the possibility of Landau and transit-time energization of the particle: we wish to focus solely on the cyclotron interaction. 
If this makes one uncomfortable, it may be justified by considering low ion beta $\beta_i = 8\pi n_i T_i/B_0^2$, where such effects (for the ions) are typically relatively weak since the typical phase velocity $\vph\sim \vA \gg \vth$. \am{Note we have also assumed that the electric and magnetic fields do not vary in the $\hat{\vx}$ direction.}
We assume that the functions $b$ and $g$ are analytic for $t$ real and that $b(y,z,\pm\infty)=g(y,z,\pm\infty)=0$.
$b$ and $g$ are related according to Faraday's law,
\begin{equation}
    \partial_t b = \vperpin \partial_z g.\label{eq:faraday}
\end{equation}
We carry out our calculation in the frame moving at $\vparin$, the parallel velocity of the ion at $t=-\infty$, and normalize according to
\am{\begin{equation}
    X= x/\rho, \quad Y= y/\rho, \quad Z=(z-\vparin t)/\rho \quad T = \Omega_{i} t,
    \label{eq:dimensionless}
\end{equation}}
where $\Omega_i=Z_i eB/m_ic$ is the ion gyrofrequency and $\rho = \vperpin/\Omega_i$ is the ion gyroradius. \am{Later, it will be useful to write $g$ and $b$ in terms of their Fourier transforms in $Y$,
\begin{align}
    g(Y,Z,T) &= \frac{1}{2\pi}\int_{-\infty}^\infty \tilde{g}(K,Z,\eta_KT)e^{iKY}\dd K,\nonumber\\
    b(Y,Z,T) &= \frac{1}{2\pi}\int_{-\infty}^\infty \tilde{b}(K,Z,\eta_KT)e^{iKY}\dd K.\label{eq:ft}
\end{align}
The dimensionless quantity $\eta_K$ appearing in the arguments of $\tilde{g}$ and $\tilde{b}$ is a bookkeeping parameter that describes how fast the fields at wavenumber $K$ vary relative to the cyclotron motion of the particle: for $\eta_K\sim1$, the fields can vary significantly over one orbit, while for $\eta_K\ll1$, they only vary a small amount. Importantly, we do not require $\eta_K\ll1$: in fact, for the main calculation that appears in Sec.~\ref{sec:main} we formally require $\varepsilon\ll\eta_K$ for all $K\gtrsim1$, i.e., $\eta_K$ cannot be too small. The case with $\eta_K\sim\varepsilon$ or smaller is dealt with in Appendix~\ref{app:slow}, where we show that our results can be extended to this case with no changes. If $\eta_K$ in some system happens to be constant with $K$, we will sometimes simply write $\eta$. Denoting $\mathrm{d}f/\mathrm{d} T=\dot{f}$,} the equations are then 
\begin{align}
    \ddot{X} &= \dot{Y},\label{eq:X}\\
    \ddot{Y} &= -\dot{X} + \varepsilon g(Y,Z,T),\label{eq:Y}\\
    \ddot{Z} &= -\varepsilon\dot{Y} b(Y,Z,T),\label{eq:Zddot}
\end{align}
which we will solve subject to the arbitrary choices for the phase of the particle $X(0)=1$, $\dot{X}(0)=0$, $Y(0)=0$, $\dot{Y}(0)=1$, $Z(0)=0$, and we have chosen the inertial frame of reference such that $\dot{Z}(-\infty)=0$.  
In the normalized variables, Faraday's law is
\begin{equation}
    \partial_T b = \partial_Z g.\label{eq:faradaynorm}
\end{equation}
Integrating (\ref{eq:X}), taking the constant of integration to be zero, and inserting the resulting equation into (\ref{eq:Y}), we have
\begin{align}
    \dot{X} &= Y,\label{eq:Xdot}\\
    \ddot{Y}+Y &= \varepsilon g(Y,Z,T). \label{eq:Yddot}
\end{align}

\section{Solution for $\varepsilon\ll\eta_K\sim1$}\label{sec:main}
We expand 
\begin{align}
X= X_0 + \varepsilon X_1 + \varepsilon^2 X_2 +\ldots\quad
Y= Y_0 + \varepsilon X_1 + \varepsilon^2 Y_2 +\ldots\quad
Z= Z_0 + \varepsilon Z_1 + \varepsilon^2 Z_2 +\ldots \label{eq:pert}
\end{align}
and proceed with our calculation.\footnote{As mentioned above, in this section we formally require that the fields do not vary too slowly at small scales: $\varepsilon\ll\eta_K\sim1$ for all $K\gtrsim1$: this does not preclude $\eta_K\ll1$, so long as $\eta_K\gg\varepsilon$. We will point out clearly where the calculation breaks down for the case of $\eta_K\sim\varepsilon\ll1$: and in Appendix~\ref{app:slow}, we will show that this can be "fixed", with no change to our final results. Note also that our orderings $\varepsilon\ll\eta_K\sim1$ and $K\sim1$ are different from the usual guiding center regime, for which $\eta_K\ll1$, $K\ll1$, and $\epsilon\sim1$.} At zeroth order in $\varepsilon$, we just have the gyration of the particle about the background field,
\begin{equation}
    X_0=-\cos{T}, \quad Y_0= \sin{T}, \quad Z_0=0,\label{eq:zeroord}
\end{equation}
according to the (arbitrary) conditions we set for $T=0$. At first order in $\varepsilon$, inserting the zeroth-order solution above for $Z_0$ and $\dot{Y}_0$ into  (\ref{eq:Yddot}) and (\ref{eq:Zddot}),
\begin{align}
    \ddot{Y}_1 + Y_1 &= g(\sin T,0,T),\label{eq:Y1ddot}\\
    \ddot{Z}_1 &= - b(\sin T,0,T)\cos(T)\label{eq:Z1ddot}.
\end{align}
Eq.~(\ref{eq:Y1ddot}) may be solved by Fourier transforming in time and back again; the solution is
\begin{equation}
    Y_1 = \int_{-\infty}^T \sin(T-T') g(\sin T',0,T')\dd T' = \dot{X}_1,
\end{equation}
and the first order $Y$-velocity is
\begin{equation}
    \dot{Y}_1 = \int_{-\infty}^T \cos(T-T')g(\sin T',0,T')\dd T'.\label{eq:Y1prim}
\end{equation}
To make further progress, we Fourier-transform in $Y$ according to Eq.~(\ref{eq:ft}), and use the identity
\begin{equation}
    e^{iK\sin T} = \sum_{n=-\infty}^\infty J_n(K) e^{inT},\label{eq:bessel}
\end{equation}
where $J_n$ are Bessel functions of the first kind. This results in
\begin{align}
    \dot{Y}_1 &= \int_{-\infty}^T \cos(T-T')\frac{1}{2\pi}\int_{-\infty}^\infty \tilde{g}(K,0,\eta_KT')\sum_{n=-\infty}^\infty J_n(K)e^{inT'}\dd K\dd T',\nonumber\\
    &= \cos T \int_{-\infty}^T \cos T'\frac{1}{2\pi}\int_{-\infty}^\infty \tilde{g}(K,0,\eta_KT')\sum_{n=-\infty}^\infty J_n(K)e^{inT'}\dd K\dd T'\nonumber\\
    &\quad+\sin T \int_{-\infty}^T \sin T'\frac{1}{2\pi}\int_{-\infty}^\infty \tilde{g}(K,0,\eta_KT')\sum_{n=-\infty}^\infty J_n(K)e^{inT'}\dd K\dd T'.\label{eq:Y1solsmall}
\end{align}
Similarly, we have
\begin{align}
    \dot{X}_1 &= \int_{-\infty}^T \sin(T-T')\frac{1}{2\pi}\int_{-\infty}^\infty \tilde{g}(K,0,\eta_KT')\sum_{n=-\infty}^\infty J_n(K)e^{inT'}\dd K\dd T',\nonumber\\
    &= \sin T \int_{-\infty}^T \cos T'\frac{1}{2\pi}\int_{-\infty}^\infty \tilde{g}(K,0,\eta_KT')\sum_{n=-\infty}^\infty J_n(K)e^{inT'}\dd K\dd T'\nonumber\\
    &\quad-\cos T \int_{-\infty}^T \sin T'\frac{1}{2\pi}\int_{-\infty}^\infty \tilde{g}(K,0,\eta_KT')\sum_{n=-\infty}^\infty J_n(K)e^{inT'}\dd K\dd T'.\label{eq:X1solsmall}
\end{align}
Combining the sinusoids and $e^{inT'}$ factors in the integrands, and using the identity $J_{n-1}(K)+J_{n+1}(K) = 2 nJ_n(K)/K$,
\begin{align}
    \dot{Y}_1 &= \frac{1}{2\pi}\cos T \int_{-\infty}^T\int_{-\infty}^\infty \tilde{g}(K,0,\eta_KT')\sum_{n=-\infty}^\infty \frac{nJ_n(K)}{K}e^{inT'}\dd K\dd T'\nonumber \\
    &\quad\,+\frac{1}{2\pi}\sin T \int_{-\infty}^T \int_{-\infty}^\infty \tilde{g}(K,0,\eta_KT')\sum_{n=-\infty}^\infty \frac{J_{n-1}(K)-J_{n+1}(K)}{2i}e^{inT'}\dd K\dd T'.\label{eq:Y1smallsol2}
\end{align}
Likewise, one finds
\begin{align}
    \dot{X}_1 &= \frac{1}{2\pi}\sin T \int_{-\infty}^T\int_{-\infty}^\infty \tilde{g}(K,0,\eta_KT')\sum_{n=-\infty}^\infty \frac{nJ_n(K)}{K}e^{inT'}\dd K\dd T'\nonumber \\
    &\quad\,-\frac{1}{2\pi}\cos T \int_{-\infty}^T \int_{-\infty}^\infty \tilde{g}(K,0,\eta_KT')\sum_{n=-\infty}^\infty \frac{J_{n-1}(K)-J_{n+1}(K)}{2i}e^{inT'}\dd K\dd T'.\label{eq:X1smallsol2}
\end{align}
Finally, using (\ref{eq:ft}) to Fourier-transform $b(Y,Z,T)$, the first-order parallel velocity is given by
\begin{equation}
    \dot{Z}_1 = -\frac{1}{2\pi} \int_{-\infty}^T \int_{-\infty}^\infty  \tilde{b}(K, 0 , \eta_K T')\sum_{n=-\infty}^\infty \frac{n J_n(K)}{K}e^{i n T'}\dd K\dd T'.\label{eq:Z1solsmall}
\end{equation}
\subsection{Change in perpendicular energy}
We are interested in the change in $v_\perp^2$ as $T\to\infty$,
\begin{equation}
    \frac{v_\perp^2}{\vperpin^2} = 1 + 2\varepsilon \left(\dot{X}_0\dot{X}_1 + \dot{Y}_0\dot{Y_1}\right)_{T\to\infty} + O(\varepsilon^2).\label{eq:dvperp2}
    \end{equation}
Using (\ref{eq:zeroord}) differentiated with respect to $T$, (\ref{eq:X1smallsol2}) and (\ref{eq:Y1smallsol2}),
\begin{align}
   \Delta = 2\left( \dot{X}_0\dot{X}_1+\dot{Y}_0\dot{Y}_1\right)_{T\to\infty} = \frac{1}{\pi}\int_{-\infty}^\infty \int_{-\infty}^\infty \tilde{g}(K,0,\eta_K T') \sum_{n=-\infty}^\infty \frac{nJ_n(K)}{K} e^{inT'}\dd K\dd T'.\label{eq:v0v1} 
\end{align}
Clearly, the contribution from $n=0$ in the sum vanishes. The integral is of the form 
\begin{equation}
    \Delta = \frac{1}{\pi}\int_{-\infty}^\infty\int_{-\infty}^\infty \sum_{n=-\infty}^\infty A(n,K,\eta_KT')e^{inT'}\dd K\dd T',\label{eq:v0v1prim}
\end{equation}
which we can perform by closing the contour in the appropriate half-plane. The dominant contribution comes from the pole of $g(K,0,s)$ (say $s_*$) closest to the real axis, so that 
\begin{equation}
\Delta \sim 2i\int_{-\infty}^\infty \sum_{n=-\infty}^\infty \textrm{sgn}(n)\,\textrm{Res} [A(n, K, s), s_*]\exp\left(- \frac{|n \mathrm{Im}\left\{s_*\right\}|}{\eta_K}\right)\dd K.\label{eq:v0v1estim}
\end{equation}
Since $A(n,K,s)=0$ for $n=0$, 
if we have that $\eta_K\ll1$ for all $K$,
this is exponentially small. At higher order in $\varepsilon$, similar exponentially-small expressions occur; this is a special case of the general conservation of adiabatic invariants to all orders \citep{kruskal1958,kruskal1962}.

Moreover, if $\eta_K\ll1$ for all $K$, we need only keep the $n=\pm1$ term, since it is obviously the largest. 
Therefore, we approximate $\Delta$ as
\begin{align}
    \Delta &\approx \frac{2}{\pi} \int_{-\infty}^\infty \cos T' \int_{-\infty}^\infty \frac{J_1(K)}{K}\tilde{g}(K,0,\eta_K T')\dd K\dd T'.\nonumber\\
    &\sim c_1\int_{-\infty}^\infty \frac{J_1(K)}{K}\mathrm{Res}\left[\tilde{g}\right]\exp(-c_2/\eta_K)\dd K\label{eq:deltasmall}
\end{align}
where $c_1,c_2$ are (system-dependent) dimensionless constants of order unity and $Res[\tilde{g}]$ denotes the residue from the pole of $\tilde{g}$ closest to the real axis, and is a function of $K$.

At this point it is worth discussing when and how our solution breaks down. First, note that it is possible to have a situation where the exponential term arising from the pole is cancelled out by part of $g$: for example, if $g(Y,Z,T) = \cos (T+\phi) g'(Y,Z,T)$. 
This is resonance, and leads to the breakdown of the ordering if $g'$ is nonzero for a time $\delta T \sim O(1/\varepsilon)$. 
We will assume this is not the case, but briefly discuss it in Appendix~\ref{app:resonance}.

Second, while we have shown that the first-order change in perpendicular kinetic energy (\ref{eq:v0v1}) is exponentially small for $\eta_K\ll1$ for all $K$, the same is not true for our expressions for the perpendicular velocities (\ref{eq:Y1smallsol2}--\ref{eq:X1smallsol2}): the second integral in each case clearly has a non-zero $n=0$ term, meaning that if the fields are left on for a time $\Delta T \gtrsim \varepsilon^{-1}$, the ordering of the solution will break down due to secularly growing terms in $\dot{Y}_1$ and $\dot{X}_1$. This could be the case, for example, if the field varies so slowly that $\eta_K\sim\varepsilon$; hence our formal restriction to larger $\eta_K$ in this section. These secular terms cancel out in the expression (\ref{eq:v0v1}) for the change in kinetic energy. Unlike the case of a true resonance, this is simply a consequence of the naive perturbation method. In Appendix~\ref{app:slow}, we use the Poincar\'e–Lindstedt method to extend the calculation to the case of arbitrarily small $\eta_K$, showing that the expression (\ref{eq:v0v1}) for the change in perpendicular kinetic energy does not change. This more involved calculation is therefore perhaps of more mathematical than physical interest, but is included for completeness.

\subsection{Scale dependence}\label{sec:scaledep}
Because of the Bessel function, the contributions to $\Delta$ from different perpendicular scale $\sim 1/K$ vary with $K$. 
For small argument ($K\ll1$), $J_n(K) \approx (K/2)^n/n!$, so that the only term that survives in Eq.~(\ref{eq:Y1smallsol2}) is $n=1$, and we may replace $J_1(K)/K$ in (\ref{eq:deltasmall}) with a scale-independent factor $1/2$.
In the opposite limit of large argument $K\gg1$, the envelope of $|J_n(K)|\sim K^{-1/2}$, and so all the terms in (\ref{eq:Y1solsmall}) become small. 
However, in turbulence, $\eta_K$ is typically an increasing function of $K$, and the exponential suppression of the heating will be less effective for larger $K$: the balance between these is system-specific, depending on $\eta_K$ and the $K$-dependence of the Fourier amplitudes $\tilde{g}$.

\subsection{Scattering contours}\label{sec:scatter}
Let us for the moment assume that the electromagnetic fluctuations are from a propagating wave or superposition of waves, with a parallel phase velocity $\vph(K)$, so that $\tilde{b}=\tilde{b}(K,Z-(\vph(K)/\vperpin)T)$, and $\tilde{g}=\tilde{g}(K,Z-(\vph/\vperpin)T)$. 
The results derived in the previous sections do not require this, but it will allow us to make contact with the usual quasilinear theory of cyclotron heating. 
It may also be directly applicable to the turbulence in the solar wind, which can be highly imbalanced: dominated by outward-going Alfv\'en waves, for which $\vph = \vA$. 
More generally, this situation could potentially also apply to nonlinear solitary waves which have a single effective $\vph$ due to the nonlinearity balancing dispersion \citep{kakutani1969,kawahara1969,hasegawa1976,mjolhus1986,mallet2023}. We note that we are ignoring the possibility of purely waves with phase velocity only in the perpendicular direction, and also electrostatic waves (which to have parallel phase velocities must also have $E_z\neq 0$ so that Faraday's law is satified). 

In a frame moving at $\vref$ compared to the laboratory, to first order in $\varepsilon$, as $T\to\infty$ we have the energy change 
\begin{align}
    v_\perp^2 + (v_\parallel - \vref)^2 &= \vperpin^2 + (\vparin - \vref)^2 + 2 \varepsilon\vperpin^2\left[ \Delta + 2 \dot{Z}_1 (\vparin-\vref)/\vperpin\right].\label{eq:contour1}
\end{align}

From Faraday's law (\ref{eq:faradaynorm}) we have
\begin{equation}
    \tilde{g}= \frac{\vparin - \vph(K)}{\vperpin} \tilde{b},
\end{equation}
so that (\ref{eq:v0v1}) can be written
\begin{equation}
    \Delta = \frac{1}{\pi} \int_{-\infty}^\infty \int_{-\infty}^\infty \frac{\vparin - \vph(K)}{\vperpin}\tilde{b}(K,0,\eta_K T') \sum_{n=-\infty}^\infty \frac{nJ_n(K)}{K} e^{inT'}\dd K\dd T'.\label{eq:Deltawave}
\end{equation}
Combining this expression with Eq.~(\ref{eq:Z1solsmall}) as $T\to\infty$, we find that
\begin{equation}
    \Delta + 2 Z_1(\vparin - \vref)/\vperpin = \frac{1}{\pi}\int_{-\infty}^\infty \int_{-\infty}^\infty \frac{\vref - \vph(K)}{\vperpin}\tilde{b}(K,0,\eta_K T') \sum_{n=-\infty}^\infty \frac{nJ_n(K)}{K} e^{inT'}\dd K\dd T'.
\end{equation}
The integrand of this expression vanishes if $\vph(K)=\vref$. The LHS is the expression appearing in square brackets in Eq.~\ref{eq:contour1}. Therefore, for a propagating wave or coherent wavepacket (e.g. a soliton), diffusion occurs along the scattering contours $v_\perp^2 + (v_\parallel - \vph)^2=\text{const}$. 
This behaviour is lost if there is not a single $\vph$, as would be the case for a dispersive wavepacket where $\vph(K)$ is not constant. This is also the case in strong, balanced Alfv\'enic turbulence, where while the linear and nonlinear frequencies are statistically in critical balance, so that $\partial_t \sim \vA\partial_z \sim u_x \partial_y$, there is a broad range of effective frequencies; or equivalently a distribution of effective phase velocities with mean zero and width $\sim \vA$.

Obviously, for a particle moving at $\vparin=\vph$, the electric field is zero \am{(again, provided that there is no electrostatic wave)}, the magnetic field is stationary in time, and thus there is no change in the perpendicular or parallel energy of the particle. 
This is encoded in the fact that for a particle moving at $\vparin=\vph$, the phase of the wave is $z-\vph t = z_0 + (\vparin-\vph)t=z_0$, independent of $t$, and so $\eta_K=0$ for all $K$.

\subsection{Example}\label{sec:example}
\begin{figure}
    \centering
    \includegraphics[width=\linewidth]{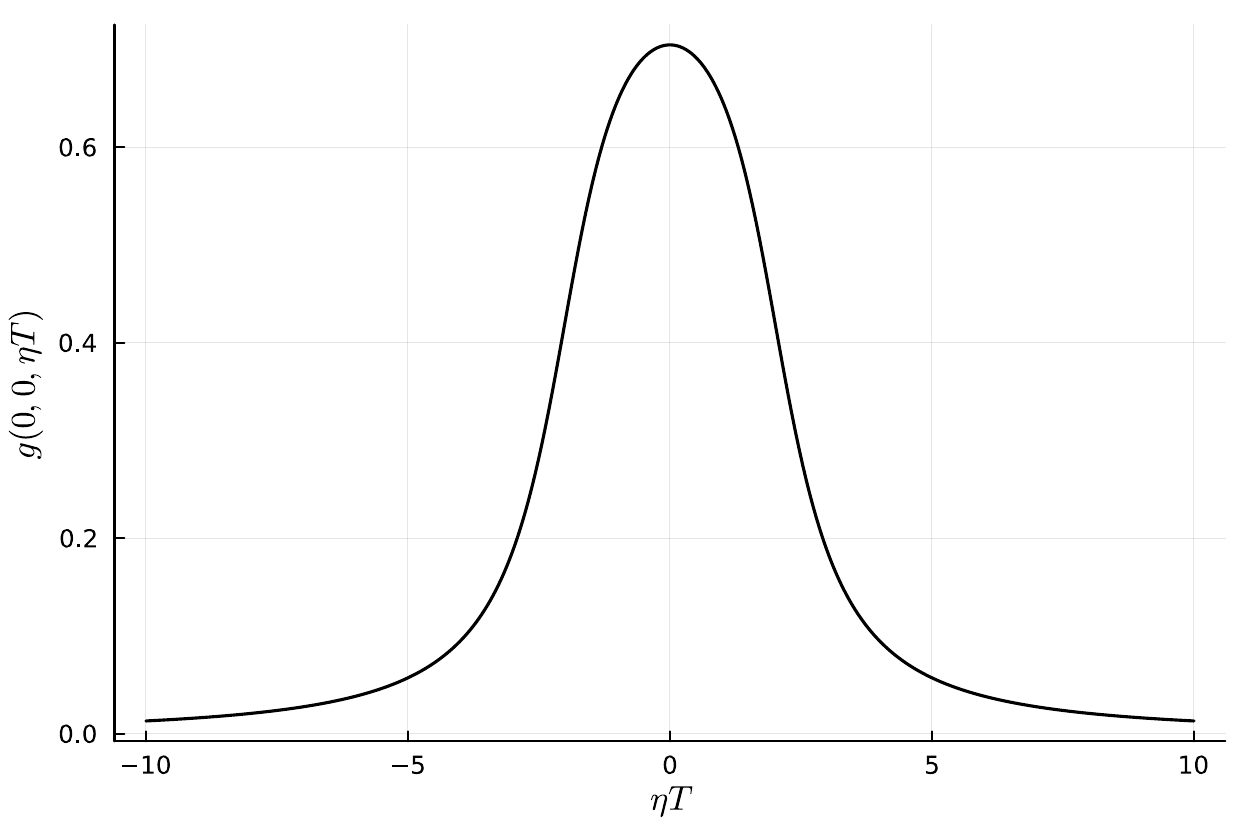}
    \caption{The functional form (\ref{eq:example}) for the time-dependence of the electric field, with $a=-2$, $b=2$.}
    \label{fig:example}
\end{figure}
Let us (for simplicity's sake) assume that $\tilde{g}(K,0,\eta_KT)=f(\eta_K T) \tilde{h}(K)$. 
As an example, we choose
\begin{equation} \label{eq:example}
    f[\eta_K T] = \frac{1}{\pi}\left\{\arctan[\eta_K(T-a)]-\arctan[\eta_K(T-b)]\right\}.
\end{equation}
We will leave the spatial pattern of the fluctuation $\tilde{h}(K)$ arbitrary since we are mainly interested in the dependence of the energy change on $\eta_K$. 
This is a model of a fluctuation that "turns on" at a rate $\eta$ around $t=a$ and then "turns off" at the same rate at $t=b$, and is plotted in Figure~\ref{fig:example}. We need to perform integrals of the form
\begin{equation}
    \int_{-\infty}^\infty e^{inT} f(\eta_K T) dT = \frac{i}{n}\int_{-\infty}^\infty e^{inT} \frac{\dd}{\dd T}f(\eta_KT) \dd T,
\end{equation}
for integer $n\neq0$, and we have integrated by parts to get the second expression\footnote{If $f(\eta\tau)$ tended to a nonzero constant at $\tau\to\pm\infty$, there would be sinusoidal terms here that can just be absorbed by the homogeneous solution for $\dot Y_1$.}. We have
\begin{equation}
    \frac{\dd}{\dd T}f(\eta_KT) = \frac{\eta_K}{\pi}\left[\frac{1}{\eta_K^2(T-a)^2+1}-\frac{1}{\eta_K^2(T-b)^2+1}\right]
\end{equation}
The poles are at $T=a\pm i/\eta$, $T=b\pm i/\eta$. We close in the appropriate half plane, and obtain for $n\neq 0$
\begin{equation}
\int_{-\infty}^\infty e^{inT} f(\eta_K T) dT = \frac{i}{n}\left[e^{ina} -e^{inb}\right]e^{-|n|/\eta_K}.
\end{equation}
Then, the perpendicular energy change (\ref{eq:v0v1}) is
\begin{align}
    \Delta= \frac{1}{\pi}\int_{-\infty}^\infty \tilde{h}(K)\sum_{\substack{n=-\infty \\ n\neq 0}}^\infty \frac{iJ_n(K)}{K} \left[e^{ina}-e^{inb}\right]e^{-|n|/\eta_K}\dd K.\label{eq:exdelta}
\end{align}

Two simplified limits are of interest: first, if $\eta_K\ll1$ and the fields vary slowly compared to the gyrofrequency, the $n=1$ terms dominate (and even they are exponentially small):
\begin{align}
    \Delta \approx \frac{2}{\pi}\left(\sin b - \sin a\right)\int_{-\infty}^\infty \tilde{h}(K)\frac{J_1(K)}{K} e^{-|n|/\eta_K}\dd K, \quad \eta_K\ll1 \,\,\forall\,\, K.
\end{align}
Second, if $\tilde{h}(K)$ only has power at small $K$, we can again drop all but the $n=1$ term, as discussed earlier in Sec.~\ref{sec:scaledep}, and additionally $J_1(K)/K\to 1/2$:
\begin{align}
    \Delta \approx \frac{1}{\pi}\left(\sin b - \sin a\right)\int_{-\infty}^\infty \tilde{h}(K) e^{-|n|/\eta_K}\dd K, \quad \tilde{h}(K)\ll1  \,\,\forall\,\, K\gtrsim1.
\end{align}

We can understand the dependence on $a$ and $b$ as depending on the phase of the particle's orbit at some reference time. 
Assuming that the ion velocity distribution is gyrotropic, each individual particle is as likely to gain or lose energy from the interaction: the average $\overline{\Delta}=0$.

\subsection{Diffusion coefficient and heating rate}\label{sec:diff}

We have so far derived an expression for the change in perpendicular energy of a single particle, $\Delta$ (Eq.~\ref{eq:v0v1}), and derived its explicit form for an example (Eq.~\ref{eq:exdelta}). Importantly, $\Delta$ can be both positive and negative, and in fact, the average over the initial gyrophase of the particle $\overline{\Delta}$ vanishes. 

Repeated interactions with coherent fluctuations will cause diffusion in energy. To be more precise, let us suppose for the moment that there are a large number of identical coherent fluctuations present, and the particle encounters one approximately every $\delta t$. Each interaction with a fluctuation occurs with a random initial gyrophase, and provides an uncorrelated kick in perpendicular kinetic energy of magnitude $\delta \kappa = m_i \vperpin^2\varepsilon \Delta/2$. This leads to an energy diffusion coefficient
\begin{equation}
    D \sim \frac{\delta \kappa^2}{\delta t}\sim\frac{1}{4}\frac{m_i^2 \vperpin^4 \varepsilon^2\overline{\Delta^2}}{\delta t},
\end{equation}
where the overline denotes averaging over the (uniform) gyrophase distribution of the particles.

If all the fluctuations are characterized by a single perpendicular scale $\lambda\sim 1/k_\perp=\rho/K$\footnote{As is common in the turbulence literature, we will interchangeably refer to "wavenumber" $k_\perp$ and "scale" $\lambda \sim 1/k_\perp$: the energy change due to a coherent structure at scale $\lambda$ is dominated by the response from  $k_\perp\sim 1/\lambda$.}, then the normalized timescale for the interaction is $\tau_\lambda\sim1/\Omega_i\eta_K$. Let us now suppose that these fluctuations are rare: in each time interval of length $\tau$, an encounter with a fluctuation occurs with probability $P$. Then, the $\delta t$ appearing in the formula for the diffusion coefficient is clearly just $\delta t = \tau/P=1/(P\Omega_i\eta_K)$, i.e.
\begin{equation}
    D \sim \frac{1}{4}\Omega_i m_i^2 \vperpin^4 P\varepsilon^2\eta_K\overline{\Delta^2}.\label{eq:DP1}
\end{equation}

Now consider the more realistic case where at each scale $\lambda$ there is an ensemble of different fluctuations, each with their own $\varepsilon$, $\overline{\Delta^2}$ and $\delta t$: we may characterize this ensemble by a joint probability distribution $P_\lambda(\varepsilon,\overline{\Delta^2},\delta t)$, noting that the arguments need not be independent. Then, we can generalize (\ref{eq:DP1}): denoting the average over the distribution of fluctuations $P_\lambda$ with angle brackets,
\begin{equation}
     D \sim\frac{\Omega_i m_i^2 \vperpin^4}{4}\left\langle \varepsilon^2\eta_K\overline{\Delta^2}\right\rangle.
\end{equation}
In the case of fluctuations that are propagating (linear or nonlinear) waves, so that $\omega = \vph k_\parallel$, the diffusion will be along the scattering contours (Sec.~\ref{sec:scatter}). 

Finally, we can use our expression Eq.~(\ref{eq:deltasmall}) for $\Delta$ to estimate
\begin{equation}
    D \sim \Omega_i m_i^2\vperpin^4 \frac{J_1^2(k_\perp\rho)}{k_\perp^2\rho^2} \varepsilon^2_{k_\perp}\eta_{k_\perp}\exp \left(-\frac{2}{\eta_{k_\perp}}\right).\label{eq:diffK}
\end{equation}
As $\eta_{k_\perp} \to 1$ from below, the diffusion becomes strong. 
To get an overall effective heating rate per unit mass, suppose $\vperpin\sim\vth$; then,
\begin{equation}
    Q_\perp \sim \frac{D(\vth)}{m_i^2\vth^2} \sim \Omega_i \vth^2 \frac{J_1^2(k_\perp\rhoth)}{k_\perp^2\rhoth^2} \varepsilon^2_{k_\perp}\eta_{k_\perp}\exp \left(-\frac{2}{\eta_{k_\perp}}\right).\label{eq:qperp}
\end{equation}
We have derived this diffusion coefficient for the energy of a single particle interacting with a distribution of fluctuations. If we consider the whole population of ions, with ion velocity distribution function $f(v_\perp)$, interacting with a single coherent fluctuation, the behaviour is also diffusive. If there is a gradient in $f(v_\perp)$, while individual particles are just as likely to gain or lose energy, the flux of particles from the region with larger $f(v_\perp)$ will be larger than the flux from the  region with smaller $f(v_\perp)$, smoothing the gradient. As an illustration, suppose that the initial distribution is uniform in gyrophase but confined to a single $\vperpin$: a ring distribution. Afterwards,
\begin{equation}
    v_\perp = \vperpin\sqrt{1+\varepsilon\Delta}\approx\vperpin(1+\varepsilon_{k_\perp}\Delta/2).
\end{equation}
The variance of this distribution is then
\begin{equation}
    \overline{ v_\perp^2} - \vperpin^2 \approx \varepsilon_{k_\perp}^2\vperpin^2\overline{\Delta^2}/4,
\end{equation}
i.e., the effective temperature has changed by an amount of order 
\begin{equation}
    \delta T_\perp \sim {m_i}\varepsilon_{k_\perp}^2\vperpin^2\overline{\Delta^2} \sim m_i \vperpin^2 \frac{J_1^2(k_\perp\rho)}{k_\perp^2\rho^2}\varepsilon_{k_\perp}^2 \exp(-2/\eta_{k_\perp}).\label{eq:deltaT}
\end{equation}
where we have used Eq.~(\ref{eq:deltasmall}) to estimate $\Delta$. As $Q_\perp\sim \delta T_\perp/\delta t$, and $\delta t\sim \eta_{k_\perp}\Omega_i$, Eqs.~(\ref{eq:qperp}) and (\ref{eq:deltaT}) agree with each other.

In the rest of the paper, we will use the theory described above to study ion heating in Alfv\'enic turbulence (using Eq.~\ref{eq:qperp}, since over a long time period each particle will interact with many fluctuations) and reconnection (using Eq.~\ref{eq:deltaT}, since the particles only interact with a reconnection exhaust once). 
To do so, it is necessary to have on hand estimates of $\varepsilon_{k_\perp}$ and $\eta_{k_\perp}$, the accuracy of the latter being more critically important: due to its presence inside the exponential cutoff, our cavalier disregard of coefficients of order unity might lead to large inaccuracies in the estimated heating rates. 
For this reason, in much of the rest of the paper we will (following the approach of \citet{chandran2010}) insert adjustable constants parameterizing these unknown coefficients in our estimates for $\varepsilon_{k_\perp}$ and $\eta_{k_\perp}$. 
Given a detailed enough knowledge of the system's dynamics, it would in principle be possible to derive these coefficients from first principles; more practically, one can fit them numerically \citep{xia2013,cerri2021,johnston2025}, although care must be taken to take account of the unrealistically limited scale separation possible in simulations of turbulence and reconnection.

There are a few different approaches to estimating $\eta_{k_\perp}$. 
In the first, we estimate $\eta_1=|\omega - k_\parallel \vparin|/\Omega_i$ where $\omega$ is some linear or nonlinear frequency of the system. 
If $\beta_i\ll1$ and we have Alfv\'enic fluctuations with $\omega \sim k_\parallel \vA\gg k_\parallel \vparin$, the $\omega$ term tends to dominate. 
In the second, we estimate $\eta_2\sim \varepsilon K$ as the (inverse of the) time it takes to $E\times B$ drift out of the structure, assuming that in reality the fields have structure in the $\hat\vx$ direction too. 
This can only possibly be relevant once $k_\perp\rhoth\gtrsim1$, since for $k_\perp\rhoth\ll1$ the fields are frozen into the plasma flow. 

Finally, one might think to estimate the time it takes for the polarization drift ($\sim \varepsilon_{k_\perp}\eta_{k_\perp}$) to cause the particle to leave the structure in the $\hat\vy$ direction \citep{mcchesney1987,chen2001,white2002}, $\eta_3 \sim \varepsilon\eta_{k_\perp} K$, where probably $\eta_{k_\perp}\sim\eta_1,\eta_2$. 
As can be seen, this is only comparable to $\eta_1$ or $\eta_2$ if $\varepsilon_{k_\perp} K \sim 1$, i.e. only at very large amplitude. 

To preview the approach of the next two sections, in Alfv\'enic turbulence (Sec.~\ref{sec:turb}) we will find that $\eta_1\sim\eta_2$, while in reconnection (Sec.~\ref{sec:rx}), we will use $\eta_2$ exclusively, assuming little structure in the parallel direction.

\section{Low-$\beta$ Alfv\'enic turbulence}\label{sec:turb}
In the Alfv\'enic turbulence present in the solar wind and corona, both $\varepsilon$ and $\eta$ depend on $k_\perp$. 
We will assume a relatively low $\beta$, so that the kinetic reduced electron heating model (KREHM) equations \citep{zocco2011} may be used; we also assume that while $k_\perp\rho_p\sim k_\perp \rho_s \sim 1$, $k_\perp d_e\ll1$, so the electrons are isothermal; where the thermal proton gyroradius $\rho_p = v_{th \rm{p}}/\Omega_p$ (different from $\rho$!), the ion sound radius is $\rho_s = \sqrt{ZT_e/m_i}/\Omega_p$, with $\Omega_p = e B/m_p c$ the proton gyrofrequency, and $d_e = c/\omega_{pe}$ is the electron inertial length. 
We want to express our results in terms of what is experimentally observable in the solar wind; namely, the $\delta B_\perp$ fluctuation amplitude as a function of the perpendicular wavenumber $k_\perp$; we will write this \markup{in velocity units as $\delta b_k = \delta B_k/\sqrt{4\pi n_{0p}m_p}$.} 
For now, we will also neglect intermittency in the turbulent fluctuation amplitude, supposing that we may characterise the amplitude at each scale by a single value $\delta b_k$. 
(The effects of intermittency will be examined in Sec.~\ref{sec:intermittency}.) 
Moreover, we assume the critical balance, such that $\omega \sim k_\perp \delta u_{e}$, where the effective electron bulk flow velocity is \am{\citep{zocco2011}}
\begin{equation}
    \boldsymbol{u}_e = \frac{c}{B_0}\hz\times \nabla_\perp \left[1 + \frac{Z}{\tau}(1-\hat{\Gamma}_0)\right]\phi,\label{eq:ue}
\end{equation}
where $\hG$ is the inverse Fourier transform of
\beq
\Gamma_0(k_\perp^2\rho_p^2/2) = I_0(k_\perp^2\rho_p^2/2)e^{-k_\perp^2\rho_p^2/2} ,
\eeq 
$I_0$ being the modified Bessel function. \am{The $\frac{Z}{\tau} (1-\hG)$ appearing in the square brackets on the RHS of (\ref{eq:ue}) results from the diamagnetic drift; the electron density fluctuations are $\delta n_e/n_{0e} = -Z/\tau (1-\hG) e\phi/T_{0e}$.} 
\am{For Alfv\'enic fluctuations, one finds that $\delta u_e \sim \alpha_k\delta b_\perp$, where 
\begin{equation}
    \alpha_k  = k_\perp \rho_p \sqrt{\frac{1}{2}\left[\frac{Z}{\tau} + \frac{1}{1-\Gamma_0(k_\perp^2\rho_p^2/2)}\right]}.\label{eq:alphak}
\end{equation}
This is true not only for linear Alfv\'en waves, but statistically even in strongly nonlinear kinetic-Alfv\'en turbulence \citep{groselj2018}: in a similar sense to the fact that $\delta u \sim \delta b$ in strong MHD turbulence \citep{marongoldreich}.} Since $\Gamma_0\approx 1-k_\perp^2\rho_p^2/2$ for $k_\perp\rho_p\ll1$, but $\Gamma_0\approx 0$ for $k_\perp \rho_p\gg1$, $\alpha_k\to 1$ as $k_\perp\rho_p\ll1$ (Alfv\'en waves) and $\alpha_k\to k_\perp\rho_{\rm th}$ for $k_\perp\rho_p\gg1$ (kinetic Alfv\'en waves). 
Thus, we have 
\begin{equation}
    \eta_{k_\perp} = \frac{\omega}{\Omega_i} \sim  \frac{2}{c_2}\frac{\alpha_k k_\perp\delta b_k}{\Omega_i}\label{eq:etaestim}
\end{equation}
where we have neglected $k_\parallel \vparin \ll \omega \sim k_\parallel \vA$ since $\beta_i$ is small, and as promised added an undetermined constant $c_2$ accounting for (several) prefactors of order unity we have neglected. 
This could in principle be corrected for the slowing down of the turbulent cascade due to dynamic alignment \citep{boldyrev,Chandran14,ms16} and/or imbalance \citep{schekochihin2022,chandran2025}. 
A more important limitation is that the KREHM equations assume that $\omega/\Omega_p\ll1$. 
We will rather flagrantly ignore this restriction in the following analysis: while it could be corrected for by using a more accurate dispersion relation, we believe it does not impact our results in a significant way. The parameter $\varepsilon_{k_\perp}=Ec/B\vperpin$ controlling the amplitude of the electric fields is similarly found\footnote{Note that due to the smallness of $\beta$ implied by our use of the KREHM equations, there is no contribution to $\phi$ from parallel magnetic field fluctuations $\delta b_\parallel$ as appears in the stochastic heating theory of \citet{cerri2021}.}: using $E\sim k_\perp\phi$ and using Eqs.~(\ref{eq:ue}) and (\ref{eq:alphak}),
\begin{equation}
    \varepsilon_{k_\perp} = \frac{Ec}{B\vperpin} \sim \frac{k_\perp^2\rho_p^2/2}{1-\Gamma_0(k_\perp^2\rho_p^2/2)} \frac{\delta b_k}{\alpha_k \vperpin}.\label{eq:epsk}
\end{equation}
The dependence of the electric field fluctuations on $k_\perp\rho_p$ is quite different from the magnetic field fluctuations. 
At $k_\perp\rho_p\gg1$, $\varepsilon_{k_\perp} \propto k_\perp\rho \delta b_k$, as can be seen in the example spectrum plotted in Fig.~\ref{fig:spectrum}. 
This means that, in the absence of dissipation, the electric field fluctuations increase with $k_\perp$ for $k_\perp\rho_p\gg1$, as observed in fully kinetic turbulence simulations \citep{groselj2018}. 
Putting this all together with the estimate for the perpendicular diffusion coefficient (\ref{eq:diffK}),
\begin{equation}
    D =  c_1m_i^2 \vperpin^2 \frac{J_1^2(k_\perp \rho)}{k_\perp^2\rho^2} \frac{k_\perp^4\rho_p^4}{\alpha_k(1-\Gamma_0(k_\perp^2\rho_p^2/2))^2} k_\perp\delta b_k^3 \exp\left(-\frac{c_2\Omega_i}{k_\perp \alpha_k\delta b_k}\right),
\end{equation}
where the order-unity constants $c_1$ and $c_2$ account for all the numerical prefactors as well as "twiddles" ($\sim$) appearing in our estimates above. 
Using Eq.~(\ref{eq:qperp}), the heating rate for thermal ions (i.e. assuming $\vperpin\sim \vth)$ is then
\begin{equation}
    Q_\perp \sim c_1 \frac{J_1^2(k_\perp \rhoth)}{k_\perp^2\rhoth^2} \frac{k_\perp^4\rho_p^4}{\alpha_k(1-\Gamma_0(k_\perp^2\rho_p^2/2))^2} k_\perp\delta b_k^3 \exp\left(-\frac{c_2\Omega_i}{k_\perp \alpha_k\delta b_k}\right).\label{eq:qperp2}
\end{equation}

For the moment assuming that the ion component of the plasma is mainly protons\footnote{This assumption is in fact implicit in Eqs.~\ref{eq:ue}--\ref{eq:epsk} earlier in this section: the minor ions have sufficiently low abundances that they do not affect the KAW dispersion relation nor polarization properties of the fluctuations.}, $\vth= v_{th \rm p}$, $\rhoth\sim\rho_p$, and $\Omega_i=\Omega_p$, we have
\begin{equation}
    Q_{\perp p} = c_1 \frac{J_1^2(k_\perp \rho_p)}{k_\perp^2\rho_p^2} \frac{k_\perp^4\rho_p^4}{\alpha_k(1-\Gamma_0(k_\perp^2\rho_p^2/2))^2} k_\perp\delta b_k^3 \exp\left(-\frac{c_2\Omega_p}{k_\perp \alpha_k\delta b_k}\right).\label{eq:qperpp}
\end{equation}
We will return to the subject of minor ions in Sec.~\ref{sec:minor}. 
Taking $k\rho_p\sim1$, we recover the expression given in \citet{chandran2010}, where the exponential suppression factor depends only on the amplitude of the fluctuations: the equivalence between exponential suppression factors based on the timescale and the amplitude for $k \rho_p\sim 1$ was first noticed by \citet{cerri2021}. 
However, we can now assess the scale dependence of the heating directly. 
It is interesting to look at this in different limits. For $k_\perp\rho_p\ll1$, we have $1-\Gamma_0\approx k_\perp^2\rho_p^2/2$, $\alpha_k \sim 1$, and $J_1(k_\perp\rho_p)\sim k_\perp\rho_p/2$. 
Then,
\begin{equation}
    Q_{\perp p} \sim c_1 k_\perp \delta b_k^3 \exp\left(-\frac{c_2\Omega_p}{k_\perp\delta b_k}\right), \quad k_\perp\rho_p\ll1.\label{eq:qperplargescale}
\end{equation}
For $k_\perp\rho_p\gg1$, $\Gamma_0\approx 0$, $\alpha_k\sim k_\perp\rho_p$ (ignoring dependence on $Z/\tau$ for simplicity), and the envelope of $|J_1(k_\perp\rho_p)|\sim (k_\perp\rho_p)^{-1/2}$, so that
\begin{equation}
    Q_{\perp p} \sim c_1 k_\perp \delta b_k^3 \exp\left(-\frac{c_2 \Omega_p}{k_\perp^2\rho_p\delta b_k}\right), \quad k_\perp\rho_p\gg1\label{eq:qperpsmallscale}
\end{equation}
The only difference is in the exponential suppression factor; the speedup in the frequency for small-scale fluctuations means we get an extra factor of $k_\perp\rho_p$ there. 
Were $\delta b_k$ independent of scale, the heating rate becomes monotonically larger towards smaller scale, since the frequency increases. Eqs. (\ref{eq:qperpp}--\ref{eq:qperpsmallscale}) can be compared with the heating rate (due to Landau/transit-time damping) in Alfv\'enic turbulence within the gyrokinetic model \citep{quataert1998,howes2010}, $Q_\perp \sim k_\perp \delta b_k^3 \gamma_p/\omega$: since $\gamma_p/\omega$ is rather small at low $\beta$, magnetic moment breaking may be the dominant heating mechanism despite the exponential suppression factor. Future work will assess this statement in a more satisfactory manner.

\subsection{Balanced turbulence}

\begin{figure}
    \centering
    \includegraphics[width=\linewidth]{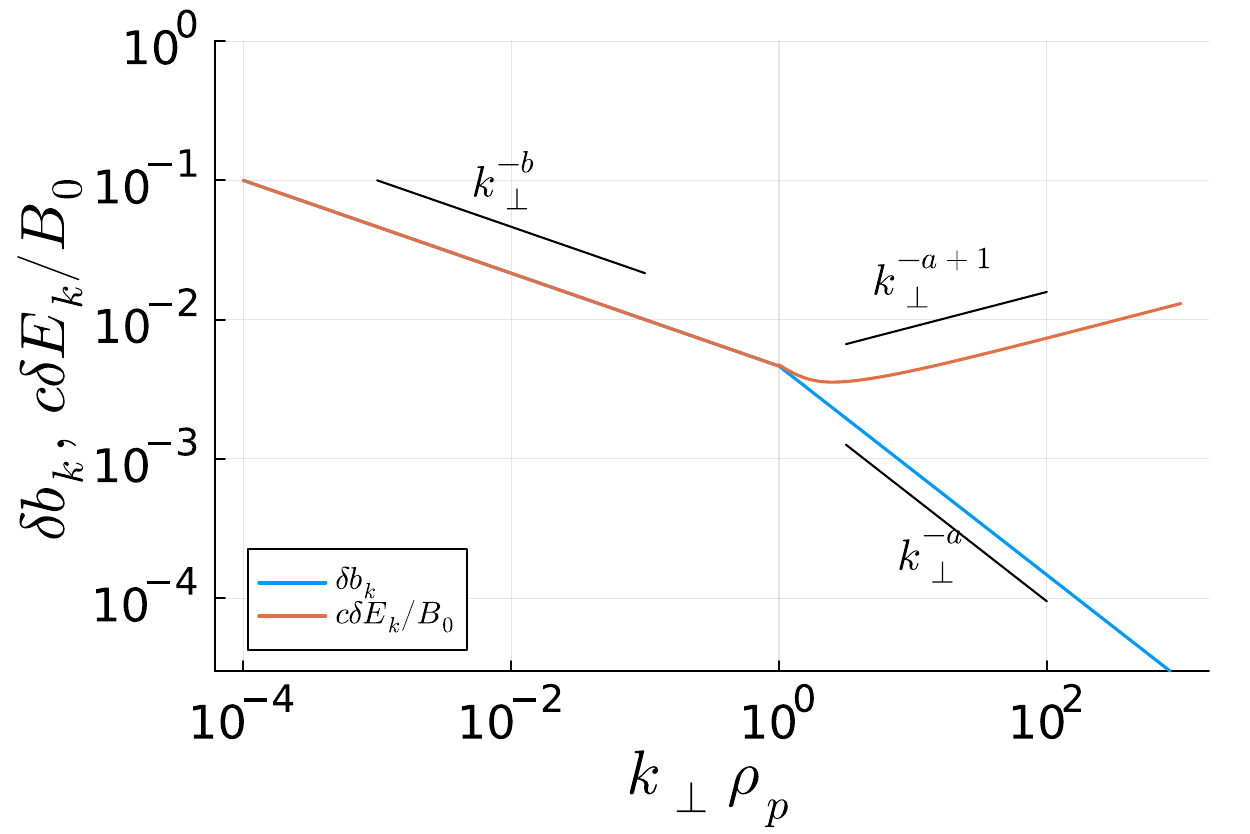}
    \caption{Typical scalings for the magnetic and electric field fluctuation amplitudes as a function of $k_\perp\rho_p$, in the absence of strong dissipation. 
    Here we have set $b=1/3$ and $a=3/4$, and the sub-ion-scale range is unrealistically long: in reality it would be cut off at the smaller of $k_\perp \sim 1/d_e$ or $k_\perp \sim 1/\rho_e$. 
    We do not model electron-scale effects in this paper.}
    \label{fig:spectrum}
\end{figure}

We will first examine the case of "balanced" turbulence, where the fluxes of Alfv\'enic fluctuations propagating parallel and antiparallel to the magnetic field are comparable: the imbalanced case is rather different, and is discussed in Sec.~\ref{sec:imbalanced} later. Typical scalings for the fluctuation amplitudes $\delta b_k$ and $c\delta E_k/B_0$ in balanced turbulence are plotted in Figure~\ref{fig:spectrum}. 
\am{At small scales, $k_\perp\rho_p\gg1$}, we have $\delta b_k \sim k_\perp^{-2/3}$ or steeper \citep{schektome2009,boldyrevkaw2012,zhou2023}; say $\delta b_k \sim \delta b_* (k_\perp\rho_p)^{-a}$, where $\delta b_*$ is then the amplitude at $k_\perp\rho_p = 1$. 
Then,
\begin{equation}
    Q_{\perp p} \sim (k_\perp\rho_p)^{-3a+1}\frac{\delta b_*^3}{\rhoth} \exp\left(-\frac{c_2 \Omega_p}{k_\perp^{2-a}\rho_p^{1-a}\delta b_*}\right), \quad k_\perp\rho_p\gg1
\end{equation}
This reaches a maximum when 
\begin{equation}
    \frac{c_2\Omega_p}{k_\perp^{2-a}\rho_p^{1-a}\delta b_*} = \frac{3a-1}{2-a},
\end{equation}
or at
\begin{equation}
    k_{max}\rho_p = \left(\frac{2-a}{3a-1}\frac{c_2\Omega_p}{\omega_*}\right)^{1/(2-a)},
\end{equation}
where $\omega_* = \delta b_*/\rho_p$ is the frequency of gyroscale fluctuations.
The maximum proton heating rate therefore occurs when $\eta_{k_\perp} = \omega/\Omega_p \sim 1$ (in an order-of-magnitude sense). Here, we remind the reader that in reality, this estimate will become inaccurate since our estimates will fail around $k_\perp d_e \sim 1$, where the spectrum steepens again \citep{stawarz2019} and our estimates for $\alpha_k$ also break down \citep{adkins2024}.

\begin{figure}
    \centering
    \includegraphics[width=\linewidth]{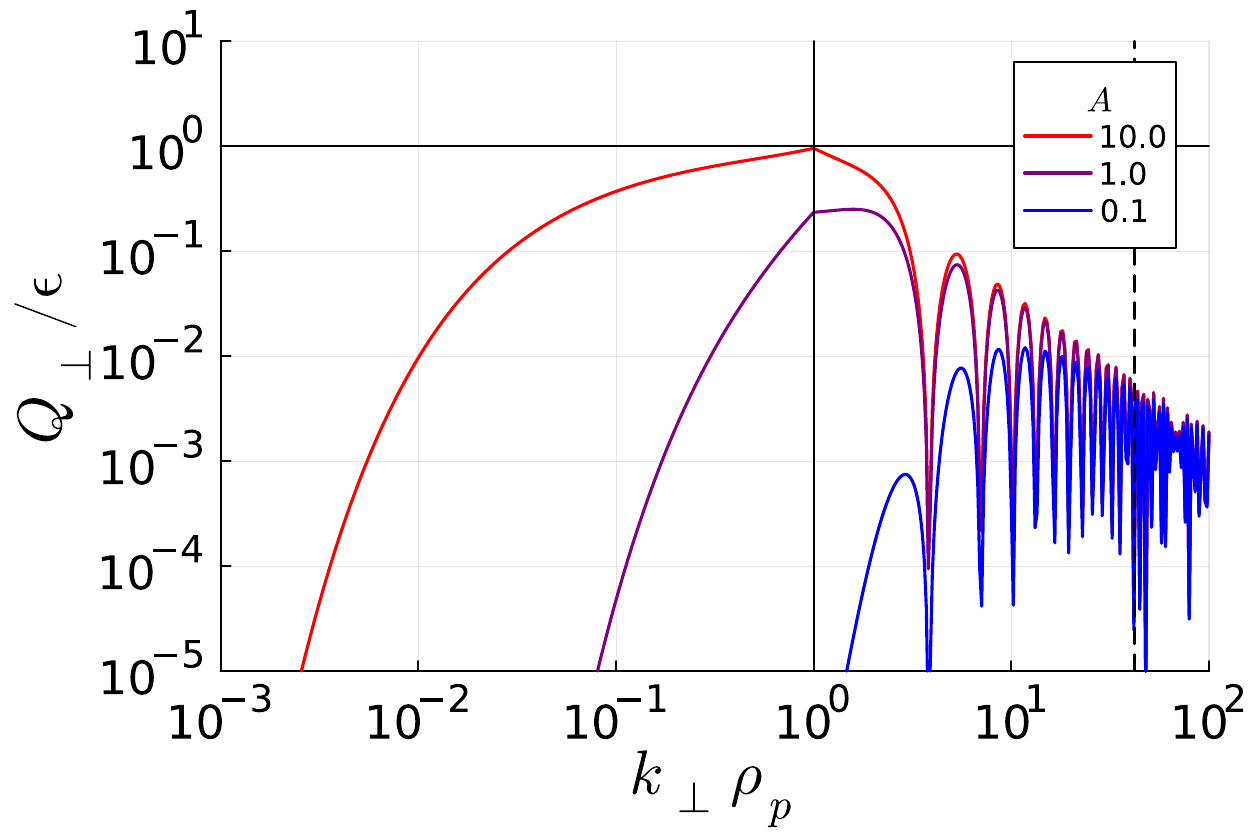}
    \caption{The proton heating rate $Q_{\perp p}$ normalized to the turbulent energy flux through scales $\epsilon=\delta b_L^3/L$, for different values of the normalized outer-scale amplitude $A$, with $L/\rho_p = 10^{4}$ and $v_{th \rm p}/\vA=0.1$. 
    The horizontal black line denotes $Q_{\perp p}/\epsilon=1$, complete damping of the turbulent cascade: in reality, if the heating approaches this line the power-law behaviour of the spectra and the constancy of $\epsilon$ will no longer be accurate. 
    The vertical solid black line denotes $k_\perp\rho_p=1$, and the vertical dashed line denotes $k_\perp\rho_e=1$: the small heating rates at or beyond the electron scales in our model (which neglects electron-scale physics) are an overestimate due to the much steeper spectrum in this range.}
    \label{fig:qperpvsK}
\end{figure}

\am{At large scales, $k_\perp\rho_p\ll1$},  the magnetic fluctuations scale as a power law, between $\delta b_k \propto k_\perp^{-1/3}$ \citep{gs95,chenmallet} and $\delta b_k\propto k_\perp^{-1/4}$ \citep{boldyrev,mallet3d,chen2020}, where the exact scaling likely depends on the regime of turbulence. 
Then, $Q_{\perp p}$ is an increasing function of $k_\perp$. 
In balanced turbulence, there is typically a smooth join between the $k_\perp\rho_p\ll1$ and $k_\perp\rho_p\gg1$ spectra, and so we expect the heating rate as a function of $k_\perp$ to reach a maximum at relatively small scales, when $\omega \sim \Omega_p$. 
This agrees (qualitatively at least) with the peak of the heating rate observed in the hybrid-kinetic numerical simulations of \citet{arzamasskiy2019} and \citet{cerri2021}. 
We have plotted the proton heating rate (\ref{eq:qperpp}) as a function of (inverse) scale $k$ in Figure~\ref{fig:qperpvsK}, for several different outer scale amplitudes $A=\delta b_L/\vA$, assuming that $L/\rho_p = 10^{4}$ and $v_{th \rm p}/\vA=0.1$. The energy flux into the turbulent  cascade is defined as $\epsilon = \delta b_L^3/L$.
The trend agrees with our analysis above: if the amplitude is high enough that $Q_{\perp p}/\epsilon \sim 1$ at $k_\perp\rho_p\sim1$, the peak heating is at the proton gyroradius scale (e.g. the red line in Fig.~\ref{fig:qperpvsK}). At lower amplitudes, the heating is less efficient, and peaks at a smaller scale (e.g. the blue line in Fig.~\ref{fig:qperpvsK}). 
The oscillations in $Q_{\perp p}$ when $k_\perp \rho_p > 1$ are due to the Bessel function; in reality, structures will have contributions from a range of $k_\perp \sim 1/\lambda$ and this behaviour will be smoothed out.

\subsection{Proton heating fraction in balanced turbulence}
The maximum proton heating rate is
\begin{equation}
    Q_{\perp max,p}\sim \frac{\delta b_*^3}{\rho_p} \left(\frac{\omega_*}{\Omega_p}\right)^{s},\label{eq:qperpmax}
\end{equation}
ignoring prefactors of order unity, and with $s=(3a-1)/(2-a)$. 
Now suppose the turbulent cascade (in the absence of dissipation) had a constant energy flux through scale $\epsilon$; by a Kolmogorov-style argument, dimensionally (as a reminder, we are neglecting intermittency in the distribution of fluctuation amplitudes),
\begin{equation}
    \epsilon\sim\delta b_*^3/\rho_p.
\end{equation} 
If $\omega_*/\Omega_p\ll1$, $Q_{\perp max}/\epsilon\ll1$ and there is no significant ion heating; the energy must be dissipated at smaller scales onto the electrons. 
Writing $\omega_* \sim \delta b_*/\rho_p$,
\begin{equation}
    \frac{Q_{\perp max,p}}{\epsilon} \sim \left(\frac{\delta b_*}{v_{th\rm p}}\right)^{s}.
\end{equation}
With $a=3/4$ (a not unreasonable value given the observed spectrum, e.g. \citet{chenkaw2010}), $s=1$, and this agrees with the expression for the ion heating fraction given in \citet{matthaeus2016}. 
It may be more useful to write this in terms of the amplitude at the outer scale $L$ of the turbulence, parametrized as
\begin{equation}
    \delta b_L = A \vA,
\end{equation}
where $\vA$ is the Alfv\'en velocity. 
Assuming $\delta b_k \propto k_\perp^{-b}$ for $k_\perp\rho_i\ll1$ (with $b=1/3$ corresponding to \citet{gs95} and $b=1/4$ corresponding to the \citet{boldyrev} including dynamic alignment), we have
\begin{equation}
    \delta b_* \sim A \vA \left(\frac{\rho_p}{L}\right)^{b},
\end{equation}
and
\begin{equation}
    \frac{Q_{\perp max,p}}{\epsilon} \sim A^{s}\beta_p^{-s/2}\left(\frac{\rho_p}{L}\right)^{sb},
\end{equation}
where the proton plasma beta $\beta_p = v_{th\rm p}^2/\vA^2$. 
Inserting $a=3/4$ ($s=1$) for simplicity,
\begin{equation}
    \frac{Q_{\perp max,p}}{\epsilon} \sim A\beta^{-1/2}\left(\frac{\rho_p}{L}\right)^{b}.
\end{equation}
\am{This estimate can be compared with other mechanisms, for example with the approach outlined in \citet{howes2024}: for a dissipation mechanism to be important, we require the turbulent system to pass a threshold in parameter space beyond which $Q_{\perp max}/\epsilon\sim 1$. Here, this threshold is described in terms of $\beta$ and $\rho_p/L$. For perpendicular ion heating to be important, we need (taking $b=1/4$)
\begin{equation}
    \frac{\rho_p}{L} \gtrsim A^{-4} \beta^{2},
\end{equation}
which is the same dependence on $\beta$ as found by \citet{howes2024} for stochastic heating \citep{chandran2010}.} Typically, $\rho_p\ll L$: for example, in the solar wind, $\rho_p/L \sim 10^{-4}$, while in the interstellar medium (ISM), $\rho_p/L \sim 10^{-11}$. 
Thus, this simple estimate suggests that ion heating should be negligible in the ISM, and account for only $5-10\%$ of the energy budget in the $\beta\sim1$ solar wind, in stark contrast to the available evidence \citep{cranmer2009}. 
This suggests that we need to incorporate additional physics into our model: in Sec.~\ref{sec:intermittency}, we show that intermittency \citep{mallet2019} results in much higher ion heating fractions, because fluctuations attain larger amplitudes such that $\omega_*/\Omega \sim 1$, leading to $Q_{\perp max,p} \sim \epsilon$.

We have neglected all heating apart from that at $k_{max}$: while amending this might increase the heating rates slightly, in practice, because of the exponential suppression of heating from low-frequency fluctuations, $Q_{\perp p}$ is quite sharply peaked at $k_{max}$.

\subsection{Forced imbalanced Alfv\'enic turbulence and the helicity barrier}\label{sec:imbalanced}

\citet{meyrand2021} found that in numerical simulations of forced imbalanced turbulence, the helicity barrier causes energy to build up at scales larger than $\rho_p$ 
with a spectral break to a very steep ($\delta b_k\propto k_\perp^{-3/2}$ or steeper) spectrum \markup{beyond $k_h\sim 2 (1-\sigma_c)^{1/4}/\rho_p\lesssim 1/\rho_p$} in the "transition range", before returning to a shallower spectrum at around $k_\perp\rho_p\sim 1$. 
We can study this situation with our heating rate expression (\ref{eq:qperplargescale}) also. 
Writing $\delta b_k \sim \delta b_h (k_\perp/k_h)^{-3/2}$, and taking $k_\perp\rho_p\lesssim1$, the frequency $\omega \sim k_\perp\delta b_k\propto k_\perp^{-1/2}$ in the transition range, a decreasing function of $k$. 
Inserting into our expression for $Q_{\perp p}$ for $k_\perp\rho_p\ll1$, we find
\begin{equation}
  Q_{\perp p} \sim k_h \delta b_h^3 \left(\frac{k_\perp}{k_h}\right)^{-7/2}   \exp\left( - \frac{c_2 \Omega_p}{k_h \delta b_h (k_\perp/k_h)^{-1/2}}\right),
\end{equation}
which decreases with increasing $k_\perp$ in the transition range. 
The assumption $\delta b_k\propto k_\perp^{-3/2}$ could be weakened; $Q_\perp$ always decreases with $k$ so long as $\delta b_k \propto k_\perp^{-1}$ or steeper. 
Thus, the maximum heating occurs at $k_h$, and is given by
\begin{equation}
    Q_{\perp max} \sim k_h \delta b_h^3 \exp\left(-\frac{c_2\Omega}{k_h\delta b_h}\right).\label{eq:hbqmax}
\end{equation}
If $k_h\delta b_h\ll\Omega$, then $Q_{\perp max}/\epsilon \sim \exp(-c_2\Omega / k_h\delta b_h)\ll1$. 
If energy is being injected into the turbulent cascade at large scales, this situation cannot be in steady state due to the presence of the helicity barrier, and thus $\delta b_h$ (and the whole large-scale spectrum) must be growing in time. 
A steady state can be achieved once $Q_{\perp max}\sim \epsilon \sim k_h\delta b_h$, which happens when $c_2\Omega / k_h\delta b_h \sim 1$, or roughly $k_h\delta b_h \sim \Omega$. 
Thus, in forced imbalanced turbulence, essentially all the energy flux from the turbulent cascade goes into ion heating. 
A more refined analysis, taking into account the electron inertia effects entering at $d_e$, shows that in fact there is a critical level of imbalance below which the helicity barrier will not form \citep{adkins2024}: we assume that in the systems we are interested in (for example, the solar corona), $1\gg \beta_e\gg m_e/m_i$ and the turbulence is extremely imbalanced, such that we can ignore this effect.

This picture of ion heating "switching on" due to the helicity barrier is, as we mentioned above, not new 
, and already predicted in \citet{meyrand2021} and \citet{squire2021}. 
Even more similar to this work, \citet{johnston2025} found using test particle simulations that ion heating in imbalanced turbulence depends on an phenomenological exponential suppression factor, controlled by the fluctuation amplitude at the scale at which the maximum frequency is reached, i.e. the transition-range break scale $k_h$ above. 

Physical systems of imbalanced turbulence, for example the solar wind, are not in fact homogeneous nor forced in the usual way, and so the helicity barrier may not have sufficient time to reach steady state and cause as sharp a transition to ion heating as suggested on the basis of forced numerical simulations \citep{meyrand2021,squire2021}. The scaling for the maximum heating rate (\ref{eq:hbqmax}) still applies, since the helicity barrier does still cause a steep transition range at scales larger than $\rho_p$.

\subsection{Intermittency}\label{sec:intermittency}
So far, we have assumed that the turbulence is characterized by a single amplitude at each scale, $\delta b_k$. 
In reality, $\delta b_k$ is a random variable at each scale, typically with a heavy large-amplitude tail; both in solar wind observations \citep{salem2009,Zhdankin12,sioulas2022} and in numerical simulations \citep{rcb,mallet3d,zhdankin16,zhdankin16intcy}. 
This can dramatically increase the ion heating fraction \citep{mallet2019,cerri2021}.

As an (extreme) toy example, suppose that the turbulence were characterized by fluctuations of a fixed amplitude $\delta b$, filling a certain scale-dependent fraction $f_k=(k_\perp L)^{-d}$ of the volume at scale $1/k_\perp$. 
In other words, the distribution of fluctuation amplitudes is 
\begin{equation}
    \delta b_k = \begin{cases}
        \delta b &\text{with probability } (k_\perp L)^{-d},\\
        0 &\text{with probability } 1-(k_\perp L)^{-d}.
    \end{cases}
\end{equation}
Requiring the energy flux $\epsilon = \langle k_\perp \delta b_k^3\rangle = k_\perp f_k\delta b^3$ to be independent of $k$, we have $d=1$. Then, the root-mean-square value of the fluctuation amplitude measured at scale $1/k_\perp$ is
\begin{equation}
    \delta b_{rms, k} = \sqrt{\langle\delta b_k^2 \rangle} = \delta b (k_\perp L)^{-1/2}.
\end{equation}
Meanwhile, the overall heating rate at large scales, given by (\ref{eq:qperplargescale}), is
\begin{equation}
    \langle Q_{\perp p}\rangle/\epsilon \sim c_1 \exp\left(-\frac{c_2\Omega_p}{k_\perp\delta b}\right),
\end{equation}
where the average is over the distribution of fluctuation amplitudes. 
Writing this solely in terms of $\delta b_{rms,k}$,
\begin{equation}
    \langle Q_{\perp p} \rangle/\epsilon \sim c_1 \exp\left(-\frac{c_2\Omega_p (k_\perp L)^{-1/2}}{k_\perp \delta b_{rms,k}}\right)
\end{equation}
Since $k_\perp L \gg 1$, this dramatically increases the overall heating rate for a given observed $\delta b_{rms,k}$, compared to the estimate without taking account the intermittency of $\delta b_k$. 
This is not, in fact, a realistic model of intermittent Alfv\'enic turbulence - we have included it here to show in a transparent way the difference that intermittency makes to the efficiency of ion heating.

It is worth mentioning that intermittency models including dynamic alignment (e.g. \citet{Chandran14,ms16}) do not necessarily increase the ion heating fraction, because the dynamic alignment between $\delta \vz^\pm$ increases the nonlinear timescale of the turbulent structures, decreasing the effectiveness of the heating due to the exponential suppression factor. 
This decrease in ion heating fraction due to dynamic alignment is the opposite to what was found earlier in \citet{mallet2019}, who used an exponential suppression factor based solely on the amplitude \citep{chandran2010}, rather than the rate of change of the fluctuations, thus finding that dynamic alignment acted to increase the stochastic heating rate.

A more promising intermittency model in this regard (which nevertheless obtains the same $-3/2$ perpendicular spectrum) is the reflection-driven turbulence model in \citet{chandran2025}, in which dynamic alignment does not play a role. 
There, larger-amplitude fluctuations naturally have higher frequencies, as required to make the ion heating more effective. 
Another approach, independent of any particular intermittency model, would be to use \emph{in situ} measurements of the distribution of turbulent fluctuations to calculate the heating rate $\langle Q_\perp\rangle$ directly. 
Parker Solar Probe has recently \am{started to explore the plasma environment very close to the sun \citep{kasper2021} where ion heating is thought to be particularly important \citep{kasper2019b}}
, and it will be interesting to assess the ion heating in this newly-explored regime.

\subsection{Minor ions}\label{sec:minor}

Observationally, as mentioned in the introduction, minor ions appear to be heated even more strongly than the protons. 
Returning to Eq.~(\ref{eq:qperp2}), and writing the ion gyrofrequency in terms of the proton gyrofrequency, $\Omega_i = Z_im_p/m_i \Omega_p$,
\begin{equation}
Q_\perp \sim c_1 \frac{J_1^2(k_\perp \rhoth)}{k_\perp^2\rhoth^2} \frac{k_\perp^4\rho_p^4}{\alpha_k(1-\Gamma_0(k_\perp^2\rho_p^2/2))^2} k_\perp\delta b_k^3 \exp\left(-\frac{c_2\Omega_p}{k_\perp \alpha_k\delta b_k}\frac{Z_im_p}{m_i}\right).
\end{equation}
Since $Z_i m_p/m_i < 1$, the exponential suppression is less effective for minor ions, exponentially increasing the heating rate relative to the protons. This is similar to what happens in the stochastic heating model applied to minor ions \citep{chandran2010c}. Note that $\rho_{th} = \rho_p (v_{th i}/v_{th p})(Z_i m_p/m_i)$, so that the gyroradius of the ions is typically larger than that of the protons: however, in the case of imbalanced turbulence where the cascade is already cut off due to the helicity barrier (Sec.~\ref{sec:imbalanced}), this may have less of an effect. An exception is in forced homogeneous numerical simulations, where the saturated state of the turbulence has such a high amplitude that fluctuations attain the minor ion's gyrofrequency at large scales, causing extreme heating of minor species \citep{zhang2025}.

\section{Reconnection}\label{sec:rx}
\begin{figure}
    \centering
    \includegraphics[width=\linewidth]{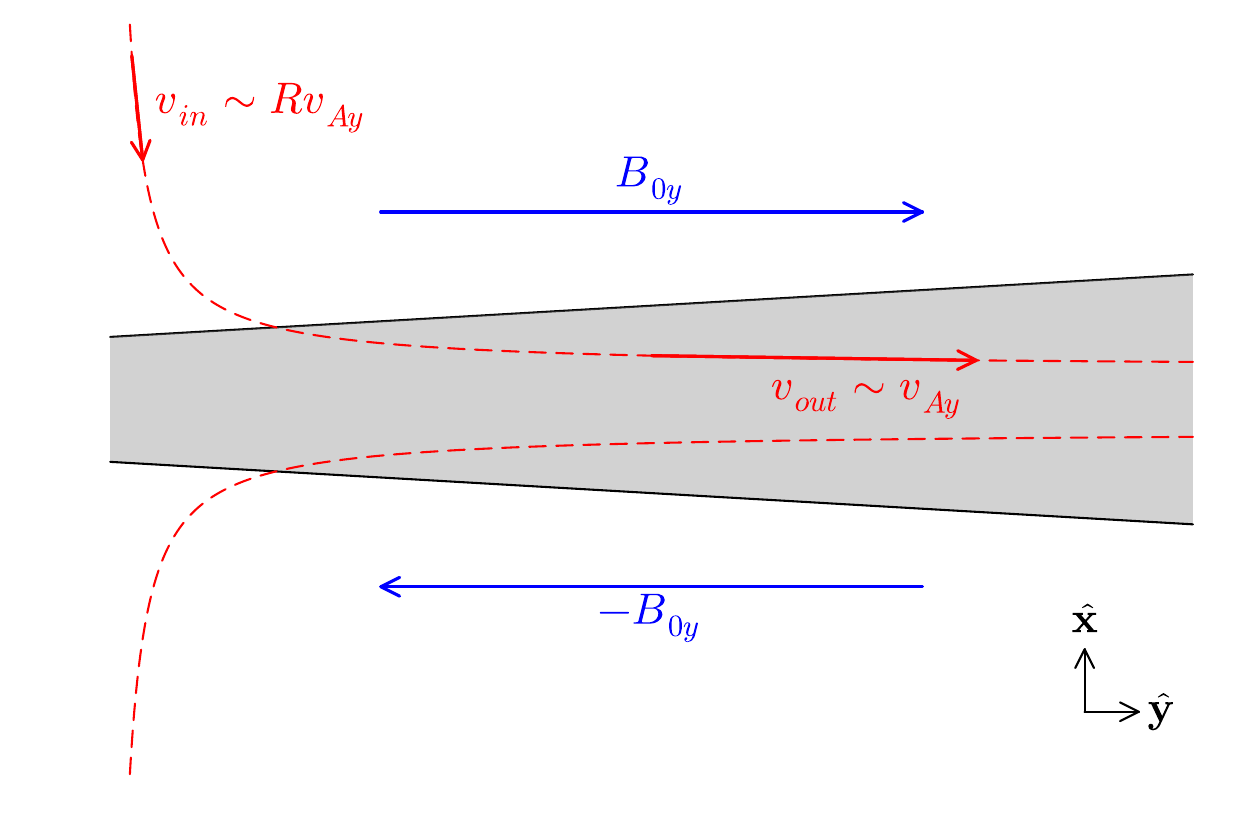}
    \caption{A crude schematic of an ion-scale reconnection exhaust. 
    The reconnecting field $\sim B_{0y}$ (blue) reverses across the exhaust. 
    An ion enters the exhaust with a slow drift velocity (red) $v_{in}\sim R v_{Ay}$ with the reconnection rate $R\sim 0.1$ and $v_{Ay} = B_{0y}/\sqrt{4\pi n_p m_p}$. 
    Within the exhaust, due to the strong electric field $E_x\sim c v_{Ay}/B_0$, where $B_0$ is the guide field, the ion takes up a drift at the Alfv\'enic outflow velocity $v_{out}\sim v_{Ay}$. 
    If this process happens in a time comparable to the ion's gyroperiod, the magnetic moment is not conserved and strong heating occurs.}
    \label{fig:rx}
\end{figure}
\citet{drake2009b} developed a theory of perpendicular ion heating in reconnection, and in \citet{drake2009a} showed that for guide-field reconnection, there was a threshold for strong ion heating by different ion species, namely
\begin{equation}
    \frac{m_i/m_p}{Z_i} \sim \frac{1}{2\pi}R^{-1}\beta^{1/2}\frac{v_{Ay}}{\vA},\label{eq:drakethreshold}
\end{equation}
where $R\sim 0.1$ is the normalized reconnection rate $u_{in}/u_{out}$, $v_{Ay}$ is the Alfv\'en speed based on the reconnecting field $\delta B_y$, and $\vA$ is the Alfv\'en speed based on the guide-field $B_0$. 
This was derived in the following way. Given a current sheet width of order $\rho_s= c_s/\Omega_p$, an inflow speed $u_{in}\sim R v_{Ay}$, the transit time of an ion from the inflow out of the sheet is of order $\tau \sim \rho_s/u_{in}$. 
Comparing this with the ion's gyrofrequency $\Omega_i=\Omega_p Z_i m_p/m_i$ and requiring $\tau\Omega_i<1$ gives (\ref{eq:drakethreshold}). 
Below the threshold, \am{the ions conserve the first adiabatic invariant $\mu$, while above the threshold, $\mu$ is not conserved} and strong ion heating occurs. 
There is an obvious equivalence between the threshold in $\eta_K$ in our estimated heating rate (\ref{eq:qperp}) and the \citet{drake2009a} theory. 
To make this more concrete, we note that in the current sheet, there is an $E_x$ driving the Alfv\'enic exhaust,
\begin{equation}
    E_x \sim \frac{B_0 v_{Ay}}{c},
\end{equation}
from which we estimate $\varepsilon$,
\begin{equation}
    \varepsilon \sim \frac{v_{Ay}}{\vth}.
\end{equation}
For $\eta\sim\eta_K\sim\text{const.}$ \am{(we assume this does not vary with $K$ since the outflow is coherent in time)}, we use the same argument as \citet{drake2009b}, leading to
\begin{equation}
    \eta \sim \frac{1}{\tau\Omega_i} \sim 2\pi R\beta^{-1/2}\frac{m_i/m_p}{Z_i} \frac{v_{Ay}}{\vA}.
\end{equation}
\am{Applying Eq.~(\ref{eq:deltaT})}, and estimating $k_\perp\rho_i\sim k_\perp\rho_s\sim 1$ so that the Bessel function term is of order unity, 
\begin{equation}
    \Delta T_\perp \sim m_i v_{Ay}^2 \exp(-\frac{2}{\eta}).
\end{equation}
Apart from numerical prefactors of order unity which we have not calculated, this agrees with Eq. 6 of \citet{drake2009b}, both in the threshold $\eta \sim 1$ and in the order-of-magnitude of the saturated total heating when the threshold is attained. 
It would also technically be possible to calculate the heating at a reconnection event using our model by specifying the functional form of the electromagnetic fields precisely, arriving at a more accurate estimate; we will leave this for future work.

This section amounts in some ways to a rederivation of the \citet{drake2009a} model. 
However, using our approach, it is perhaps more obvious why the acceleration of the ion into the outflow $E\times B$ velocity is accompanied by ion heating. 
\citet{drake2009b} suggest that "the reflected particles interpenetrate with particles that have already crossed the boundary of the exhaust but have not passed
through the reversal region", thus gaining an effective temperature. 
In our approach it is obvious that the diffusion is due to the random initial phase of the particle as it enters the reversal region of the exhaust. 
While the energy change of each ion $\Delta$ is, on average, zero, considering the plasma distribution function as a whole, the ions acquire a broad range of energies.

\section{Discussion}
We have analysed the interaction of an ion with a localized, coherent fluctuation, deriving the change in the ion's perpendicular and parallel kinetic energy. 
To lowest order, the energy change is described by Eq.~(\ref{eq:echange}), which depends linearly on the amplitude of the fluctuating fields and on a factor $\exp(-1/\eta)$, where $\eta\sim1/\tau\Omega_i$, with $\tau$ the characteristic timescale of the fluctuation and $\Omega_i$ the ion's gyrofrequency. 
For the whole population of ions, the interaction leads to diffusion in energy and heating. 
This leads to weak heating for $\eta\ll1$, and strong heating for $\eta\sim 1$: this reflects the well-known conservation of the magnetic moment for $\eta\ll1$, and is similar to previous theories of stochastic heating \citep{chandran2010}. 
As part of our derivation, we have recovered the fact that, if the fluctuation is a wavepacket with a fixed phase velocity $\vph$, the diffusion is along circular scattering contours in $v_\parallel$-$v_\perp$ space, centered on $\vph$. 
(Sec.~\ref{sec:scatter}), a result usually associated with quasilinear cyclotron resonant heating \citep{kennel1966}. 
Thus, our results combine the physics of stochastic heating and cyclotron heating in a single theoretical framework, based on the interaction of ions with coherent structures. 
In many systems, our approach based on individual localized fluctuations may be more appropriate than the assumption of infinite plane waves usually required to derive the results of quasilinear theory, for example in strongly nonlinear turbulence and reconnection. 
The model is also quite easy to apply to different physical situations: one needs estimates for a characteristic timescale relative to the gyroperiod, $1/\eta$, as well as an estimate of the fluctuation amplitude $\varepsilon$ of the structure.

We have applied our results to low-$\beta$ Alfv\'enic turbulence, obtaining a simple expression for the fraction of the turbulent flux absorbed by the protons \am{in both balanced turbulence and in imbalanced turbulence with a helicity barrier \citep{meyrand2021}, as may be present in the fast Alfv\'enic solar wind \citep{bowen2020,mcintyre2024}. Our theoretical estimate for the proton heating fraction compares well with the numerically observed estimate for the ion heating fraction in \citet{matthaeus2016}.} 
We also show that our model is well-suited to incorporating different models of intermittency, which we predict should enhance the ion heating: one could also use the observational data to assess this enhancement directly. 
Moreover, we have determined how the perpendicular lengthscale of the fluctuation affects the heating rates: at smaller amplitudes, the heating does not peak at $k_\perp\rhoth\sim 1$, but at a smaller scale, as was previously observed in the hybrid-kinetic simulations of \citet{arzamasskiy2019}. 
We also show that the heating of minor ions is greatly enhanced over the proton heating, another property associated with both cyclotron \citep{kasper2013} and stochastic \citep{chandran2010c} heating. 
Our framework can also model ion heating in reconnection, naturally producing perpendicular heating with a threshold similar to the theory and numerical simulations of \citet{drake2009a,drake2009b}. 
The similarity of our results for heating due to reconnection and turbulence suggest that the disparity between these two paradigms for coronal heating \citep{klimchuk2015,chandran2009} may not be as drastic as often thought.

\citet{johnston2025} have shown within the framework of quasilinear theory that it is possible to recover an exponential or exponential-like suppression of heating, similar to our model and to the original \citet{chandran2010} stochastic heating theory. 
Moreover, they show that test-particle heating in both balanced and imbalanced turbulence simulations seems to be well described by such an exponential suppression factor. In both \citet{johnston2025} and this paper, the exponential factor depends on the typical frequency or inverse timescale of the interactions compared to the ion cyclotron frequency, i.e. $\eta\sim 1/\tau\Omega_i$, reinforcing the fact that the relevant physics in both cases is the breaking of the magnetic moment conservation. However, the source of the suppression factor is different: in \citet{johnston2025}, the suppression factor in imbalanced turbulence is due to the steep $k_\parallel$ spectrum outside the critical balance cone leading to very little power at the high resonant frequency,
whereas in our case, it comes from the fact that $\mu$ is an adiabatic invariant as $\eta\to0$, conserved to all orders.

\textbf{Declaration of Interests.--} The authors report no conflict of interest.

\textbf{Acknowledgements.--} 
AM was supported by NASA grants 80NSSC21K0462 and 80NSSC21K1766. BC was supported in part by NASA under grant number 80NSSC24K0171.
KGK was partially supported by NASA grant 80NSSC24K0171 and contract 80ARC021C0001. TE acknowledges funding from The Chuck Lorre Family Foundation Big Bang Theory Graduate Fellowship and NASA grant 80NSSC20K1285. All authors also acknowledge support from NASA contract NNN06AA01C.
AM thanks J. Squire, Z. Johnston, and M. Kunz for many useful conversations, and the three referees for insightful comments, which significantly improved this paper.

\appendix

\section{Drifts}\label{app:drifts}
As well as calculating the behaviour of the velocity as $T\to\infty$, we might be interested in the drift velocity of the particle at finite~$T$. It is easy to see that only the terms with $n=0$ in (\ref{eq:X1solsmall}) and (\ref{eq:Y1solsmall}) contribute to drifts.
Integrating the $n=0$ term of (\ref{eq:X1solsmall}) by parts, 
\begin{align}
    \dot X_{1d} &= \frac{1}{2\pi}\sin T \left\{\left[ \sin T \int_{-\infty}^\infty J_0(K)g(K,0,\eta_K T)\mathrm{d}K\right]_{-\infty}^T \right.\nonumber\\
    &\left.\quad\quad\quad\quad\quad\quad\quad-\int_{-\infty}^T\sin T' \frac{\mathrm{d}}{\mathrm{d}T'}\int_{-\infty}^\infty J_0(K)g(K,0,\eta_K T)\mathrm{d}K \mathrm{d}T\right\}\nonumber\\
    &\quad-\frac{1}{2\pi}\cos T \left\{-\left[ \cos T \int_{-\infty}^\infty J_0(K)g(K,0,\eta_K T){\mathrm{d}} K\right]_{-\infty}^T 
    \right.\nonumber\\
    &\left.\quad\quad\quad\quad\quad\quad\quad+\int_{-\infty}^T\cos T' \frac{\mathrm{d}}{\mathrm{d}T'}\int_{-\infty}^\infty J_0(K)g(K,0,\eta_K T)\mathrm{d}K \mathrm{d}T\right\}\nonumber \\
    &= \frac{1}{2\pi}\int_{-\infty}^\infty J_0(K)g(K,0,\eta_K T)\mathrm{d}K - \cos T\int_{-\infty}^T\cos T' \frac{\mathrm{d}}{\mathrm{d}T'}\int_{-\infty}^\infty J_0(K)g(K,0,\eta_K T)\mathrm{d}K \mathrm{d}T \nonumber\\&\quad \quad \quad \quad \quad \quad-\sin T\int_{-\infty}^T\sin T' \frac{\mathrm{d}}{\mathrm{d}T'}g(0,0,\eta_K T') \mathrm{d}T.
\end{align}
The first term is the gyroaveraged $E\times B$ drift velocity. 
Repeating the integration by parts,
\begin{align}
    \dot X_{1d} &= \frac{1}{2\pi}\int_{-\infty}^\infty J_0(K)g(K,0,\eta_K T)\mathrm{d}K\nonumber\\ &\quad- \frac{1}{2\pi}\cos T \left\{ 
    \left[\sin T \frac{\mathrm{d}}{\mathrm{d}T'}\int_{-\infty}^\infty J_0(K)g(K,0,\eta_K T)\mathrm{d}K\right]_{-\infty}^T \nonumber\right.\\
    &\left.\quad\quad\quad\quad\quad\quad- \int_{-\infty}^T \sin T' \frac{\mathrm{d}^2}{\mathrm{d}T'^2} \int_{-\infty}^\infty J_0(K)g(K,0,\eta_K T)\mathrm{d}K\mathrm{d}T' 
    \right\}
    \nonumber\\
    &\quad+ \frac{1}{2\pi}\sin T \left\{
    \left[\cos T \frac{\mathrm{d}}{\mathrm{d}T'}\int_{-\infty}^\infty J_0(K)g(K,0,\eta_K T)\mathrm{d}K\right]_{-\infty}^T \nonumber\right.\\
    &\left.\quad\quad\quad\quad\quad\quad- \int_{-\infty}^T \cos T' \frac{\mathrm{d}^2}{\mathrm{d}T'^2} \int_{-\infty}^\infty J_0(K)g(K,0,\eta_K T)\mathrm{d}K
    \mathrm{d}T'
    \right\}
    \nonumber\\
    &= \frac{1}{2\pi}\int_{-\infty}^\infty J_0(K)g(K,0,\eta_K T)\mathrm{d}K \nonumber\\&\quad\quad+ \frac{1}{2\pi}\cos T \int_{-\infty}^T \sin T'\frac{\mathrm{d}^2}{\mathrm{d}T'^2} \int_{-\infty}^\infty J_0(K)g(K,0,\eta_K T)\mathrm{d}K\mathrm{d}T' \nonumber\\& \quad \quad- \frac{1}{2\pi}\sin T\int_{-\infty}^T \cos T'\frac{\mathrm{d}^2}{\mathrm{d}T'^2} \int_{-\infty}^\infty J_0(K)g(K,0,\eta_K T)\mathrm{d}K\mathrm{d}T'.  
\end{align}
Comparing this and (\ref{eq:X1solsmall}), it is easy to see the pattern:
\begin{align}
    \dot X_{1d} &= \frac{1}{2\pi}\sum_{m=0}^M (-1)^m\frac{\mathrm{d}^{2m}}{\mathrm{d}T^{2m}}\int_{-\infty}^\infty J_0(K)g(K,0,\eta_K T)\mathrm{d}K \nonumber\\
    &\quad+ \frac{(-1)^{M+1}}{2\pi}\cos T \int_{-\infty}^T\sin T' \frac{\mathrm{d}^{2M+2}}{\mathrm{d}T'^{2M+2}}\int_{-\infty}^\infty J_0(K)g(K,0,\eta_K T)\mathrm{d}K \mathrm{d}T'\nonumber\\
    &\quad - \frac{(-1)^{M+1}}{2\pi}\sin T \int_{-\infty}^T\cos T' \frac{\mathrm{d}^{2M+2}}{\mathrm{d}T'^{2M+2}}\int_{-\infty}^\infty J_0(K)g(K,0,\eta_K T)\mathrm{d}K \mathrm{d}T',
\end{align}
a series of corrections to the lowest-order gyroaveraged $\vE\times\vB$ velocity \citep{stephens2017}. 
Taking $M\to\infty$, we obtain the infinite series
\begin{equation}
    \dot X_{1d} = \frac{1}{2\pi}\sum_{m=0}^\infty (-1)^m\frac{\mathrm{d}^{2m}}{\mathrm{d}T^{2m}}\int_{-\infty}^\infty J_0(K)\tilde{g}(K,0,\eta_K T)\mathrm{d}K.
\end{equation}
Differentiating, we obtain a similar series for the drifts in $\hat{\vy}$ direction,
\begin{equation}
    \dot Y_{1d} = \frac{1}{2\pi}\sum_{m=0}^\infty (-1)^m\frac{\mathrm{d}^{2m+1}}{\mathrm{d}T^{2m+1}}\int_{-\infty}^\infty J_0(K)\tilde{g}(K,0,\eta_K T)\mathrm{d}K,
\end{equation}
with the first term being the gyroaveraged polarization drift. 
For $\eta\ll1$, each successive term in the sum is smaller than the next by a factor $\sim\eta^2$. 
If $\eta\sim 1$, drifts at all orders are comparable and the series does not converge. As $T\to\infty$ the drift part of the motion vanishes since we required $g(y,z,\infty)=0$.

\section{Resonance}\label{app:resonance}
We could modify the example in Sec.~\ref{sec:example} by multiplying by a sinusoid,

\begin{equation}
    \tilde{g}(K,0,\eta_K T) = \frac{1}{\pi}\tilde{h}(K)\cos\left(\nu t + \phi\right)\left[\arctan(\eta_K(T-a))-\arctan(\eta_K(T-b))\right].
\end{equation}

Expanding the products of cosines and sines appearing in the integral solution for $\Delta$ (\ref{eq:v0v1}), we find that there are terms in the integrands multiplied by $\exp(\pm i(\nu\pm1)t)$. For $\nu = \pm1$, secular terms therefore appear in the solution for $\dot Y_1$, scaling with the time interval over which the sinusoidally-varying fields are applied. 
If $b-a \gtrsim 1/\varepsilon$, then $\dot Y_1\gtrsim 1$ in this resonant case. 
However, in practice, this type of fluctuation clearly involves an overall rate of change of the fields $\eta' \sim 1$, so that $\exp(-1/\eta')\sim 1$ and in terms of scalings, the possibility of this resonant type of fluctuation makes little difference to the end result. 
In the rest of the paper, we ignore this subtlety, and assume that our fluctuation is not resonant.

\section{The case with $\eta_K\sim\varepsilon$}\label{app:slow}
Examining our expressions for $\dot{X}_1$ and $\dot{Y}_1$ (\ref{eq:Y1smallsol2}--\ref{eq:X1smallsol2}), it is possible to see that in fact, the coefficient with $n=0$ resulting from the second line does not vanish. As mentioned in the main paper, this means that if the fluctuating field is applied over a time longer than $O(1/\varepsilon)$, our ordering in $\varepsilon$ breaks down, since this secular term will become large. This is not a true resonance, however, and results from the fact that our expansion procedure does not take into account the nearly periodic nature of the system.

To illustrate the issue, we first consider a simplified example. Suppose that the electric field is given by $g(Y,Z,T) = Y$, constant in time but depending on $Y$. Then, our equation for $Y$ is
\begin{equation}
    \ddot Y + Y = \epsilon Y.
\end{equation}
This has the exact solution $Y=\sin[(1-\varepsilon)T]$, i.e., the variation of the electric field in space has introduced a frequency shift \citep{stephens2017}. Now suppose we instead attempted to expand in $\varepsilon$: we would instead get $Y_0=\sin T$, and at first order,
\begin{equation}
    \ddot{Y}_1 + Y_1 = Y_0 =\sin T,
\end{equation}
which is secular. To avoid this, we can use the Poincar\'e-Lindstedt method \citep{bender2013}, introducing a stretched time variable $\tau=pT$, with
\begin{equation}
    p = 1 + \varepsilon p_1 + \varepsilon^2 p_2 +\ldots
    \label{eq:defpex}
\end{equation}
Now, at first order, we have
\begin{align}
    \ddot{Y}_1 +Y_1 &= Y_0-2p_1\ddot{Y}_0,\\
    \ddot{Y}_1 + Y_1 &= (1+2p_1)\sin T,
\end{align}
where we have, in an abuse of notation, redefined the overdot to mean differentiation by $\tau$. The solution that avoids a secular term is $p_1 = -1/2$, which agrees with the exact solution up to first order in $\varepsilon$.

Our problem unfortunately does not have an readily available exact solution, but we can still apply the Poincar\'e-Lindstedt method. We order $\eta_K\sim \varepsilon$. Because of the time dependence of the electromagnetic fields, the stretching of time variable $\tau=pT$ now itself depends slowly on time, with
\begin{equation}
    p = 1 + \varepsilon p_1(\varepsilon T) + \varepsilon^2 p_2 (\varepsilon T) +\ldots
    \label{eq:defp}
\end{equation}
Our differential equation for $Y_1$ (replacing \ref{eq:Y1ddot}) becomes
\begin{equation}
    \ddot{Y}_1 + Y_1 = g(\sin \tau, 0 , T) -2p_1\sin \tau.
\end{equation}
This may be solved (as before) by Fourier transforming in time and back again, and the solution for the velocity is
\begin{equation}
    \dot{Y}_1 = \int_{-\infty}^\tau \cos(\tau-\tau')\left[g(\sin\tau',0,\tau') + 2p_1\sin\tau'\right]\mathrm{d}\tau'.\label{eq:Y1primstretch}
\end{equation}
Following the same procedure as in Sec.~\ref{sec:main}, Eqs.~(\ref{eq:Y1prim}--\ref{eq:Y1smallsol2}) (Fourier transforming in $Y$ according to (\ref{eq:ft}), applying the Bessel function identity (\ref{eq:bessel}), and expanding the cosine and combining the resulting sinusoidal terms in the integrands), we obtain
\begin{align}
    \dot{Y}_1 &= \frac{1}{2\pi}\cos\tau \int_{-\infty}^\tau\int_{-\infty}^\infty g(K,0,\eta_K\tau')\sum_{n=-\infty}^\infty \frac{nJ_n(K)}{K}e^{in\tau'}dK + p_1\sin2\tau'd\tau'\nonumber \\
    &\quad\,+\frac{1}{2\pi}\sin\tau \int_{-\infty}^\tau \int_{-\infty}^\infty g(K,0,\eta_K\tau')\sum_{n=-\infty}^\infty \frac{J_{n-1}(K)-J_{n+1}(K)}{2i}e^{in\tau'}dK\nonumber\\ &\quad\quad\quad\quad\quad\quad\quad\quad\quad\quad\quad+ p_1(1-\cos2\tau')d\tau'.\label{eq:Y1stretch}
\end{align}
A secular term would result if there were a non-zero term in the $\tau'$ integrand of (\ref{eq:Y1stretch}) with no sinusoidal variation - as discussed above, such a term would break the ordering of our solution if the electric field $g$ remains "on" for a time of order $1/\varepsilon$. 
In the second integral of (\ref{eq:Y1stretch}), one such term results from the $n=0$ term, while there is another in the $p_1$ term. 
To ensure they cancel, we choose
\begin{equation}
    p_1 = -i\int_{-\infty}^\infty J_1(K) \tilde{g}(K,0,\eta_K\tau) dK = \int_0^\infty J_1(K) \mathrm{Im}\{\tilde{g}(K,0,\eta_K\tau)\} dK\label{eq:p1}
\end{equation}
where the second equality follows since the electric field is real; thus, $p_1$ is real. 
Because $J_1(K)\approx K/2$ for $K\ll1$, if $g$ varies only on large scales, 
\begin{equation}
    p_1 \approx -\left.\frac{1}{2}\frac{dg}{dY}\right|_{Y,Z=0}.\label{eq:p1largescale}
\end{equation}
This agrees with our simplified example with $g=Y$ above. In other words, if the electric field varies only a small amount over the gyroradius of the particle, $p_1$ will also be small (by a factor $k_\perp\rho$).

With this analysis, it is now proven that even when $\eta_K\sim \varepsilon$ or smaller, $\dot{Y}_1$ (and $\dot{X}_1$, by identical arguments) are exponentially small as $T\to\infty$, as is $\Delta$.

\bibliographystyle{jpp}
\bibliography{mainbib2024}
\end{document}